\documentclass[apj,revtex4]{emulateapj}
\usepackage{textcomp}
\usepackage{epsfig, epstopdf}
\usepackage{longtable}
\usepackage{lscape}
\usepackage{amssymb,amsmath}

\newcommand{\cf}{{\ifmmode{C_{\rm f}}\else{$C_{\rm f}$}\fi}}
\newcommand{\zem}{{\ifmmode{z_{\rm em}}\else{$z_{\rm em}$}\fi}}
\newcommand{\zabs}{{\ifmmode{z_{\rm abs}}\else{$z_{\rm abs}$}\fi}}
\newcommand{\zs}{{\ifmmode{z_{\rm s}}\else{$z_{\rm s}$}\fi}}
\newcommand{\zl}{{\ifmmode{z_{\rm l}}\else{$z_{\rm l}$}\fi}}
\newcommand{\kms}{{\ifmmode{{\rm km~s}^{-1}}\else{km~s$^{-1}$}\fi}}
\newcommand{\vej}{{\ifmmode{v_{\rm ej}}\else{$v_{\rm ej}$}\fi}}
\newcommand{\vrot}{{\ifmmode{v_{\rm rot}}\else{$v_{\rm rot}$}\fi}}
\newcommand{\cm}{{\ifmmode{{\rm cm}^{-1}}\else{cm$^{-1}$}\fi}}
\newcommand{\cmm}{{\ifmmode{{\rm cm}^{-2}}\else{cm$^{-2}$}\fi}}
\newcommand{\cmmm}{{\ifmmode{{\rm cm}^{-3}}\else{cm$^{-3}$}\fi}}
\newcommand{\lya}{Ly$\alpha$} 
\newcommand{\lyb}{Ly$\beta$} 

\newcounter{species} 
\def\ion#1#2{\setcounter{species}{#2}#1$\;${\scriptsize\Roman{species}}\relax}

\slugcomment{\today}
\slugcomment{A complete version can be obtained by http://www.shinshu-u.ac.jp/faculty/general/chair/astro/arXiv/sdss1029.pdf}

\shorttitle{Multi-Sightline Observation of NALs}
\shortauthors{Misawa et al.}

\begin{document}

\title{Multi-Sightline Observation of Narrow Absorption Lines in
  Lensed Quasar SDSS~J1029+2623\altaffilmark{1,2}}

\footnotetext[1]{Based on data collected at Subaru Telescope, which is
  operated by the National Astronomical Observatory of Japan.}
\footnotetext[2]{Based on observations obtained at the European
  Southern Observatory at La Silla, Chile in programs 092.B-0512(A)}

\author{Toru Misawa\altaffilmark{3},
        Cristian Saez\altaffilmark{4,5},
        Jane C. Charlton\altaffilmark{6},
        Michael Eracleous\altaffilmark{6,7},
        George Chartas\altaffilmark{8},
        Franz E. Bauer\altaffilmark{9,10,11},
        Naohisa Inada\altaffilmark{12},
    and Hisakazu Uchiyama\altaffilmark{13}
}

\altaffiltext{3}{School of General Education, Shinshu University,
  3-1-1 Asahi, Matsumoto, Nagano 390-8621, Japan}
\altaffiltext{4}{Korea Astronomy and Space Science Institute (KASI),
  61-1, Hwaam-dong, Yuseong-gu, Deajeon 305-348, Republic of Korea}
\altaffiltext{5}{Department of Astronomy, University of Maryland,
  College Park, MD 20742-2421} \altaffiltext{6}{Department of
  Astronomy and Astrophysics, Pennsylvania State University,
  University Park, PA 16802} \altaffiltext{7}{Department of Astronomy
  \& Astrophysics and Center for Gravitational Wave Physics, The
  Pennsylvania State University, 525 Davey Lab, University Park, PA
  16802} \altaffiltext{8}{Department of Physics and Astronomy, College
  of Charleston, SC 29424} \altaffiltext{9}{Instituto de
  Astrof\'{\i}sica, Facultad de F\'{i}sica, Pontificia Universidad
  Cat\'{o}lica de Chile, Casilla 306, Santiago 22, Chile}
\altaffiltext{10}{Millennium Institute of Astrophysics (MAS), Nuncio
  Monse\~{n}or S\'{o}tero Sanz 100, Providencia, Santiago, Chile}
\altaffiltext{11}{Space Science Institute, 4750 Walnut Street, Suite
  205, Boulder, Colorado 80301}
\altaffiltext{12}{Department of Physics, National Institute of
  Technology, Nara College, Yamatokohriyama, Nara 639-1080, Japan}
\altaffiltext{13}{Department of Astronomy, School of Science, Graduate
  University for Advanced Studies, Mitaka, Tokyo 181-8588, Japan}
\email{misawatr@shinshu-u.ac.jp}

\begin{abstract}
We exploit the widely-separated images of the lensed quasar
SDSS~J1029+2623 (\zem=2.197, $\theta =22^{\prime\prime}\!\!.5$) to
observe its outflowing wind through two different sightlines. We
present an analysis of three observations, including two with the
Subaru telescope in 2010 February \citep{mis13} and 2014 April
\citep{mis14b}, separated by 4 years, and one with the Very Large
Telescope, separated from the second Subaru observation by $\sim$2
months.  We detect 66 narrow absorption lines (NALs), of which 24 are
classified as {\it intrinsic} NALs that are physically associated with
the quasar based on partial coverage analysis.  The velocities of
intrinsic NALs appear to cluster around values of \vej\ $\sim$ 59,000,
43,000, and 29,000 \kms, which is reminiscent of filamentary
structures obtained by numerical simulations.  There are no common
intrinsic NALs at the same redshift along the two sightlines, implying
that the transverse size of the NAL absorbers should be smaller than
the sightline distance between two lensed images.  In addition to the
NALs with large ejection velocities of \vej\ $>$ 1,000~\kms, we also
detect broader proximity absorption lines (PALs) at \zabs\ $\sim$
\zem. The PALs are likely to arise in outflowing gas at a distance of
$r$ $\leq$ 620~pc from the central black hole with an electron density
of $n_{\rm e}$ $\geq$ 8.7$\times$10$^{3}$~\cmmm. These limits are
based on the assumption that the variability of the lines is due to
recombination.  We discuss the implications of these results on the
three-dimensional structure of the outflow.
\end{abstract}


\section{INTRODUCTION}
AGN outflows, potentially powered by one or more of a variety of
mechanisms (e.g., radiation force, magnetic pressure, and
magnetocentrifugal force), are important ingredients of quasar central
engines and likely play a role in quasar and galaxy
formation/evolution because: 1) they extract angular momentum from
accretion disks allowing gas accretion to proceed
\citep[e.g.,][]{bla82,emm92,kon94,eve05}, leading to the growth of
black holes, 2) they also provide energy and momentum feedback to the
interstellar medium of host galaxies and to the intergalactic medium
(IGM), and inhibit star formation \citep[e.g.,][]{spr05}, and 3) they
may promote metal enrichment of the intergalactic medium (IGM)
\citep[e.g.,][]{ham97b,gab06}.  These outflowing gases, which are
difficult to observe directly, have been detected as absorption lines
in the spectra of about 50\%\ of all quasars
\citep[e.g.,][]{ves03,wis04,mis07a,nes08,muz13}. However, a limitation
of these past studies is that they observe the outflowing gas {\it
  only} along a single sightline (i.e., one dimension) toward the
nucleus of each quasar, although the absorber's physical conditions
probably depend on the location/orientation at which we observe it
\citep[e.g.,][]{gan01,elv00}.  Thus, the internal structure of
outflowing winds is still largely unknown.

Multiple quasar images, produced by gravitational lensing, provide a
unique way to study the outflowing gas along more than one
sightline. Lensed quasars with large image separation angles have a
higher chance of revealing structural differences in the outflowing
winds, especially in the vicinity of the continuum source.  In this
sense, the quasar images that are lensed by a cluster of galaxies
(rather than a single massive galaxy) are very promising targets.
Among three such lensed quasars, SDSS J1029+2623 at $z_{em}$
$\sim$2.197 \citep{ina06,ogu08} is the best target because i) it has
the largest lensed quasar image separation ($\theta$ $\sim$
22$^{\prime\prime}\!\!$.5) ever observed (see Figure~1 of
\citealt{ina06}), and ii) it exhibits absorption features in the blue
wings of the \ion{C}{4}, \ion{N}{5}, and \lya\ emission lines with
ejection velocity of \vej\ $\leq$ 1,000~\kms, which could be the
result of outflowing gas moving toward us from the central region. We
call them Proximity Absorption Lines (PALs) throughout this paper. We
define PALs as a subcategory of Narrow Absorption Lines (NALs) with
ejection velocity of \vej\ $\leq$ 1,000~\kms.  We use this terminology
throughout the paper to separate PALs from NALs with larger ejection
velocities.

\citet{mis13} obtained high-resolution spectra of the brighter two of
the lensed images (images~A and B) with the Subaru telescope in 2010
February and found several clear signs that the origin of the PALs is
indeed in the outflowing gas.  First, they show the signature of
partial coverage, which means the absorbers do not cover the
background flux source completely. There also exists a clear
difference in the absorption profiles between the spectra of images~A
and B, which can be explained by either of the following two
scenarios: (a) time variability of the absorption features over a time
scale corresponding to the time delay between the two images ({\it
  time variation scenario}; \citealt{cha07})\footnote[14]{The time
  delay between images~A and B is $\Delta t_{\rm AB}$ $\sim$ 774~days
  in the sense of A leading B, while the time delay between images~B
  and C $\Delta t_{\rm BC}$ is only a few days \citep{foh13}.}, or (b)
a difference in the absorption between the different sightlines of the
outflowing wind ({\it multi-sightline scenario};
\citealt{che03,gre06}). However, with a single-epoch observation we
cannot distinguish between these scenarios.  \citet{mis14b} performed
a second observation about four years (1514~days in the observed
frame) after the first observation (which is longer than the time
delay between images~A and B, $\Delta t_{\rm AB}$ $\sim$ 744~days),
and found that the PALs were nearly stable and that most of the
differences between images~A and B still remained.  This evidence
suggests a {\it multi-sightline scenario} where the absorber's size
should be smaller than the physical distance between the sightlines of
the lensed images, thus not covering both sightlines.  A possible
explanation is that there are a number of small {\it clumpy} clouds in
the outflowing stream.  Indeed, some of the outflowing gas is expected
to consist of small gas clouds ($d_{\rm cloud}$ $\leq$ 10$^{-3}$~pc)
with very large gas densities ($n_e$ $\geq$ 10$^6$~\cmmm;
\citealt{ham13,jos14}).  Furthermore, recent radiation-MHD simulations
by \citet{tak13} reproduce variable clumpy structures with typical
sizes of 20 times the gravitational radius ($R_{\rm g}$),
corresponding to $d_{\rm cloud}$ $\sim$ 5$\times$10$^{-4}$~pc,
assuming a black-hole mass for SDSS~J1029+2623 of $M_{\rm BH}\sim
10^{8.72}~{\rm M}_{\odot}$ \citep{mis13}.

Such clumpy clouds can be examined more easily through narrow
absorption lines (NALs, hereafter) with large offset velocities from
the quasar because the corresponding absorbers are (i) probably
smaller than the PAL absorbers and (ii) not so crowded in velocity
space as the PAL absorbers.  Indeed, there are many NALs detected in
the spectra of both images~A and B of SDSS~J1029+2623. Their origin is
not only the outflowing wind ({\it intrinsic} NALs, hereafter) but
also cosmologically intervening gas such as foreground galaxies and
the IGM ({\it intervening} NALs, hereafter).  Although it has been
traditionally believed that many NALs that fall within 5,000~\kms\ of
the quasar emission redshift (termed {\it associated} absorption lines
or AALs) are physically associated with the quasar
\citep[e.g.,][]{wey79}, we can separate intrinsic NALs from
intervening ones more effectively by performing partial coverage
analysis. In order to form a global picture of the outflowing wind, we
need to understand the physical conditions of NAL absorbers (i.e.,
highly accelerated gas) as well as PAL absorbers (i.e., weakly
accelerated gas).

In this paper, we present the results from our new spectroscopic
observation of SDSS~J1029+2623 taken with the Very Large Telescope
(VLT), which enables us for the first time to identify {\it intrinsic}
NALs in a high quality spectrum of this object.  In \S2, we describe
the observations and data reduction. The methods used for absorption
line detection and covering factor analysis are outlined in \S3. The
results and discussion are presented in \S4 and \S5. Finally, we
summarize our results in \S6.  We use a cosmology with $H_{0}$=70
\kms~Mpc$^{-1}$, $\Omega_{m}$=0.3, and $\Omega_{\Lambda}$=0.7
throughout the paper.

\section{OBSERVATIONS}
We acquired high resolution spectra of the brightest two of the three
lensed images of SDSS~J1029+2623, A and B with $V$ = 18.72 and
18.67~mags, with the VLT using the Ultraviolet and Visual Echelle
Spectrograph (UVES) in queue mode (ESO program 092.B-0512(A)).  The
observations were performed from 2014 January 28 to February 26 (epoch
E2, hereafter), which is $\sim$4~years after the first observation on
2010 February 10 (epoch E1, hereafter; \citealt{mis13}), and
$\sim$2~months before the third observation on 2014 April 4 (epoch E3,
hereafter; \citealt{mis14b}) with Subaru using the High Dispersion
Spectrograph (HDS).  We used a slit width of $1.\!\!^{\prime\prime}2$,
corresponding to $R$ $\sim$ 33,000, while \citet{mis13,mis14b} took
$R$ $\sim$ 30,000 and 36,000 spectra using Subaru/HDS.  The wavelength
coverage is 3300--6600 \AA\ in the 390/564~nm setting, which covers
the \ion{O}{6}, \ion{N}{5}, \ion{Si}{4}, and \ion{C}{4} doublets as
well as the \lya\ absorption line at \zabs\ $\sim$ \zem.  We also
adopted 2$\times$2 pixel binning in both the spatial and dispersion
directions to increase the S/N ratio.  The total integration time is
26,670s and the final S/N ratio is about 23~pix$^{-1}$ around
4700\AA\ for both images.

We reduced the data to extract the one-dimensional spectra in a
standard manner using the UVES Common Pipeline Library (CPL release
6.6).  We could not separate the third image (image~C, $V$ = 20.63)
from image~B completely because the typical seeing of our observation
($\sim$ 1.$^{\!\!\prime\prime}$0--1.$^{\!\!\prime\prime}$8) was
comparable to the separation angle between images~B and C ($\theta$
$\sim$ 1.$^{\!\!\prime\prime}$85)\footnote[15]{On the other hand, a
  flux contamination from the image~C is almost negligible in the
  image~B spectrum because the former is much fainter than the
  latter.}.

Table~\ref{t1} gives a log of the current observation with VLT/UVES as
well as our past observations with Subaru/HDS, in which we list the
target name, date of observation, telescope/instrument used, spectral
resolution, total exposure time, and signal-to-noise ratio (S/N). The
S/N is evaluated around 4700~\AA, close to the \ion{C}{4} mini-BAL. In
Figure~\ref{f1}, we show normalized spectra over the full wavelength
range of our observations for images~A and B.  These spectra were
binned every 0.5\AA\ for display purposes, and the 1$\sigma$ errors
are also shown.

\section{DATA ANALYSIS}
First, using the line detection code {\sc search}, written by Chris
Churchill, we detect all absorption features whose confidence level is
greater than 5$\sigma$ in the normalized spectrum of each lensed
image. We then identify \ion{N}{5}, \ion{C}{4}, and \ion{Si}{4}
doublets in the regions from $-$1,000~\kms\ to
5,000~\kms\ (\ion{N}{5}), from $-$1,000~\kms\ to
70,000~\kms\ (\ion{C}{4}), and from $-$1,000~\kms\ to
40,000~\kms\ (\ion{Si}{4}) around the corresponding emission lines,
with the maximum velocity set in order to avoid the
\lya\ forest\footnote[16]{Here, we define the velocity offset as
  positive for blueshifted NALs from the quasar emission redshift that
  is determined from \ion{Mg}{2} emission line \citep{ina06}.}.  We
also search for the \ion{Mg}{2} doublet in the whole range of the
spectra.  Absorption troughs that are separated by nonabsorbed regions
are considered to be separate lines.  In total, 4 \ion{Mg}{2}, 19
\ion{C}{4}, 2 \ion{N}{5}, and 7 \ion{Si}{4} doublets are identified in
the image~A spectrum, while 3 \ion{Mg}{2}, 22 \ion{C}{4}, 2
\ion{N}{5}, and 7 \ion{Si}{4} doublets are identified in the image~B
spectrum.  The equivalent widths of both blue and red members of
doublets are measured for each line by integrating across the
absorption profile and these are listed in Table~\ref{t2}.  We also
searched for 10 single metal lines (\ion{O}{1}~$\lambda$1302,
\ion{Si}{2}~$\lambda$1190, \ion{Si}{2}~$\lambda$1193,
\ion{Si}{2}~$\lambda$1260, \ion{Si}{2}~$\lambda$1527,
\ion{Al}{2}~$\lambda$1671, \ion{C}{2}~$\lambda$1036,
\ion{C}{2}~$\lambda$1335, \ion{Si}{3}~$\lambda$1207, and
\ion{C}{3}~$\lambda$1548) as well as \lya\ and \lyb\ and detected
about 200 lines at the same redshift as the doublet lines. These are
summarized in Table~\ref{t3}. Other single lines or unidentified lines
are not shown in Figure~\ref{f1} and Tables~\ref{t1}--\ref{t3} even if
they are detected at a confidence greater than 5$\sigma$.

\subsection{dN/dz Analysis}
One of the important properties of intrinsic NALs is a number density
excess of high-ionization doublets per unit redshift (i.e., dN/dz;
\citealt[][and references therein]{ham97a}).  In order to compare the
dN/dz from our spectra with those from our previous study based on 37
quasar spectra \citep{mis07a}, we construct a complete sample
including only NALs whose blue doublet members would be detected even
in the lowest S/N spectrum in \citet{mis07a}. The corresponding lower
limits of rest-frame equivalent widths (EWs) are EW$_{\rm min}^{\rm
  rest}$ = 0.056$~\AA$, 0.038$~\AA$, and 0.054$~\AA$ for \ion{C}{4},
\ion{N}{5}, and \ion{Si}{4}, respectively. Here, the values of
EW$_{\rm min}^{\rm rest}$ depend on the S/N ratio of the observed
spectrum as
\begin{equation}
 EW_{\rm min}^{\rm rest} = \frac{-U^2+U\sqrt{U^2+4({\rm S/N})^2(M_L^2
     M_c^{-1} + M_L)}}{2({\rm S/N})^2 (1+z_{abs})} \times
 \Delta\lambda (\AA),
\label{eqn:1}
\end{equation}
where $U$ is the confidence level of the EW defined as EW/$\sigma(\rm
EW)$, and $M_L$ and $M_C$ are the numbers of pixels over which the
equivalent width and the continuum level are determined
\citep{you79,tyt87}.  Using equation (1), we confirm that the S/N
of our VLT spectra is always larger than the required values
except for the region between $\lambda_{\rm obs}$ $\sim$ 4500 --
4525~$\AA$ (Figure~\ref{f2}).  After removing weak NALs with EW$^{\rm
  rest}$ $<$ EW$_{\rm min}^{\rm rest}$, we have 12 \ion{C}{4}, 2
\ion{N}{5}, and 3 \ion{Si}{4} doublets in image~A spectrum and 11
\ion{C}{4}, 2 \ion{N}{5}, and 3 \ion{Si}{4} doublets in the image~B
spectrum.  We will call this the ``homogeneous'' NAL sample,
hereafter.  Following \citet{mis07a}, we also combined multiple NALs
lying within 200~\kms\ of each other into a single NAL ``system''
because clustered lines are probably not independent even if they have
a cosmologically intervening origin \citep[e.g.,][]{sar88}.

All identified doublets (including both the homogeneous and
inhomogeneous NAL samples) are listed in Table~\ref{t2}, in which
multiple NALs within 200~\kms\ of each other are separated by
horizontal lines.  The table gives the ion name ($ion$), flux-weighted
absorption redshift (\zabs), ejection velocity (\vej) supposing they
originate in the outflow, rest frame EWs of blue and red members of
the doublet (EW$_{\rm b}^{\rm rest}$, EW$_{\rm r}^{\rm rest}$),
identification number in Figure~\ref{f1} (ID), reliability class of
intrinsic lines (described later), ionization class (described later),
and velocity difference from the first doublet at the lowest redshift
in each absorption system ($\Delta v$).  Table~\ref{t3} summarizes
other information including the flux-weighted line width of each
system on a velocity scale, $\sigma$($v$), as defined in
\citet{mis07a}, and other transitions that are detected at the same
redshift, as well as the column density ($\log N$), Doppler parameter,
$b$, and covering factor, \cf, for each absorption component in the
system.

\subsection{Partial Coverage Analysis}
Among several criteria, i) time variability, ii) partial coverage, and
iii) line locking are the most reliable properties to distinguish
intrinsic NALs from intervening NALs \citep[][and references
  therein]{bar97,ham97a}. When compared with broad intrinsic
absorption lines, intrinsic NALs are less likely to vary
\citep{mis14a,che15} and when they do vary, their variation amplitude
is small \citep{wis04,mis14a}. Therefore, the variability criterion
does not offer an efficient way of identifying intrinsic NALs.

Partial coverage analysis is quite useful for our spectra because the
resolving power is high enough to deblend NALs into multiple
components.  Using the Voigt profile fitting code {\sc minfit}
\citep{chu97,chu03}, we deblended NALs into 86 and 91 components in
images~A and B, respectively. We do not include the \ion{Si}{4} NAL at
\zabs\ = 1.8909 because the blue and red members of the doublet are
both blended with other lines.  With {\sc minfit}, we fit each NAL
profile using the redshift ($z$), column density ($\log N$ in \cmm),
Doppler parameter ($b$ in \kms), and covering factor (\cf) as free
parameters.  The covering factor (\cf) is the fraction of photons from
the background source that pass through the absorber. If the
background source is uniformly bright, then \cf\ also represents the
fraction of the background source (i.e., the continuum source and/or
broad emission line regions, BELRs) that is occulted by foreground
absorbers along our sightline.  If \cf\ is less than unity, it is
likely that the absorbers are part of a quasar outflow because
cosmologically intervening absorbers like substructures in foreground
galaxies and the IGM are less likely to have internal structures as
small as a size of the background flux sources
\citep[e.g.,][]{wam95,bar97}.  The covering factor is evaluated in an
unbiased manner as
\begin{equation}
C_{f} = \frac{(R_{r}-1)^{2}}{1+R_{b}-2R_{r}}\; ,
\label{eqn:1}
\end{equation}
where $R_{b}$ and $R_{r}$ are the residual (i.e., unabsorbed) fluxes
of the blue and red members of a doublet in the normalized spectrum
\citep[c.f.,][]{ham97b,bar97,cre99}.  If {\sc minfit} gives unphysical
covering factors for some components, e.g., negative or greater than
1, we rerun the code assuming \cf\ = 1 only for those components
because the \cf\ values are very sensitive to continuum level errors,
especially for full coverage doublets \citep{mis05}.  In addition to
the {\it fitting} method above, we evaluate \cf\ values for each {\it
  pixel} as \citet{gan99} did ({\it pixel-by-pixel} method).  The
fitting results by both methods are shown in Figure~\ref{f3} and the
fit parameters by the fitting method are summarized in Table~\ref{t3}.
We have confirmed that the \cf\ values provided by the two methods are
in good agreement with each other.

\section{RESULTS}

\subsection{Narrow Absorption Lines (NALs)}
Using our NAL sample toward the lensed images of SDSS~J1029+2623, we
perform several statistical analyses. In these analyses, in order to
avoid any possible biases we do not include the two \ion{C}{4} NALs at
\zabs\ $\sim$ 1.9322 and 1.9788 that are covered in the Subaru
spectrum \citep{mis13} but not covered by our VLT data.

\subsubsection{dN/dz analysis}
In Table~\ref{t4}, we summarize the number density of homogeneous NAL
systems per unit redshift (dN/dz) or per unit velocity offset from the
quasar (dN/d$\beta$) for \ion{C}{4}, \ion{N}{4}, and \ion{Si}{4},
along with the Poisson noise in these quantities \citep{geh86}.  All
systems are also classified into one of two categories according to
the offset velocity from the quasar; associated absorption lines
(AALs) with \vej\ $\leq$ 5,000~\kms, and non-AALs with \vej\ $>$
5,000~\kms, following \citet{mis07a}.  We found that the dN/dz values
for \ion{C}{4}, \ion{N}{5}, and \ion{Si}{4} toward this one quasar,
SDSS~J1029+2623, are larger than the average values in the larger
sample of \citet{mis07a} by factors of $\sim$3, $\sim$6, and $\sim$2,
respectively, although this enhancement is not statistically
significant because of the small number of NALs in our spectra.
Because \citet{mis07a} discovered that at least $\sim$20\%\ of
\ion{C}{4} NALs originate from winds based on partial coverage
analysis, statistically we expect to detect two or more intrinsic
\ion{C}{4} NALs among $\sim$10 homogeneous \ion{C}{4} NALs in the
spectra of images~A and B.

\subsubsection{Intrinsic or Intervening NALs}
High-velocity NALs blueshifted with \vej\ $>$ 1000~\kms\ are expected
to be outside of the range of velocities where BELs are found, and
thus, they absorb mostly continuum light. The existence of partial
coverage in high-velocity NALs suggests that the size of absorbers is
comparable to or smaller than the continuum source (i.e., $d_{\rm
  cloud}$ $\leq$ $R_{\rm cont}$).  Following \citet{mis07a}, we
separate all NALs (including those in the inhomogeneous sample) into
three classes (classes-A, B, and C) based on partial coverage
analysis, where class-A includes NALs most reliably classified as
intrinsic while class-C includes NALs that are consistent with full
coverage or that cannot be classified. Class-B contains NALs that show
line-locking, which is a signature of a radiatively driven outflowing
wind and is only detectable if our sightline is approximately parallel
to the gas motion, as often seen in NALs \citep[e.g.,][]{ben05,bow14}.
Class-B also contains systems that have tentative evidence for partial
coverage.  As a result, 4 \ion{C}{4} NALs (including PALs) are
classified as intrinsic (two class-A and two class-B) among the 12
\ion{C}{4} NALs in the homogeneous sample from the spectrum of
image~A, while 3 \ion{C}{4} NALs are classified as intrinsic (two
class-A and one class-B) among 11 NALs in the spectrum of image~B.
The fraction of intrinsic NALs (27 -- 33\%) is somewhat larger than
the average value of $\sim$20\%\ \citep{mis07a}, although our sample
size is small.  If we include the inhomogeneous sample, 5 \ion{C}{4}
NALs are classified as 3 class-A and 2 class-B among 19 NALs toward
image~A, while 9 out of 22 \ion{C}{4} NALs are classified into 4
class-A and 5 class-B NALs toward image~B.
In addition to intrinsic \ion{C}{4} NALs, we detected two class-B
\ion{Si}{4} NALs only toward image~A in homogeneous sample, and three
class-B \ion{Si}{4} NALs in each of the image~A and B spectra after
including systems from the inhomogeneous sample.
It is also noteworthy that we detect a large number of line-locked
NALs: 2 \ion{C}{4} and 3 \ion{Si}{4} NALs in image~A and 5 \ion{C}{4}
and 3 \ion{Si}{4} NALs in image~B, while only five systems are
line-locked among 138 homogeneous NALs toward 37 quasars
\citep{mis07a}.  This result strongly suggests that our sightline
toward the central source is almost parallel to the outflowing
streamline.

\subsubsection{Ionization Conditions}
The outflowing winds in the vicinity of the flux source are probably
more highly ionized than most intervening absorbers due to strong UV
radiation from the continuum source, although it depends on the gas
density of the absorbers. Indeed, a high ionization state has been
used as one indicator of the intrinsic properties of absorbers
\citep[][and references therein]{ham97a}.  Broad absorption lines
(BALs) are often classified into three categories according to their
ionization level: high-ionization BALs (HiBALs), low-ionization BALs
(LoBALs), and extremely low-ionization BALs showing \ion{Fe}{2} lines
(FeLoBALs) \citep{wey91}.  A similar classification has also been
performed for NALs \citep{ber94,mis07a}.  Motivated by the literature,
we classify our NALs into three categories, based on the detection of
absorption lines in low (ionization potential; IP $<$ 25~eV),
intermediate (IP = 35 -- 50~eV), and high (IP $>$ 60~eV) ionization
levels\footnote[17]{An important caveat here is that we do not
  necessarily detect all absorption lines because of the detection
  limits due to line strength, observed wavelength coverage, and line
  blending (e.g., in \lya\ forest).}.  Interestingly, some class-A NAL
systems include low ionization lines (see Table~\ref{t2}), which
suggests that intrinsic NAL absorbers have multiple phases with
different ionization states, as noted in \citet{mis07a}.

\subsubsection{Similarities in NALs between Sightlines}
Here, we present a new method of identifying intrinsic NALs by
exploiting our multi-sightline observation. If NALs with similar line
profiles are detected at the same redshift toward two sightlines, the
corresponding absorber must have a size larger than the physical
distance between the two sightlines. The absorber also cannot have any
internal structures on the scale of the sightline separation. Although
foreground galaxies and IGM structures also can cover both sight lines
(whose physical separation is $\sim$kpc or $\sim$Mpc scale), they
usually have significant internal velocity structures on those scales,
as often seen in the spectra of lensed quasars \citep[e.g.,][]{ell04}.
Only intrinsic absorbers can satisfy the requirement of having very
similar profiles because their sizes can be larger than the sightline
separation (e.g., sub-parsec scale) if they are at a small distance
from the flux source (e.g., $r$ $<$ 1~kpc).  Based on the ejection
velocity distribution of class-A, B, and C NALs (Figure~\ref{f4}), in
Figure~\ref{f5} we summarize the distribution of velocity differences
between NALs and PALs in the two sightlines ($\Delta v$) for systems
within $\Delta v$ $\leq$ 200~\kms\ between the two sightlines.  The
distribution is almost uniform up to $\Delta v$ $\sim$ 200~\kms\ with
a clear peak near $\Delta v$ $\sim$ 0~\kms.  Among eight NAL pairs
with $\Delta v$ $\leq$ 10~\kms, four are \ion{C}{4} and \ion{N}{5}
PALs at \zabs\ $\sim$ \zem, which we will discuss later.  The other
four NAL pairs are \ion{C}{4} NALs at \zabs\ $\sim$ 1.8909, 2.1349,
and 2.1819, and a \ion{Si}{4} pair at \zabs\ $\sim$ 1.8909.  Although
the velocity shift is very small for these NAL pairs, their line
profiles are clearly different as compared in Figure~\ref{f6}.  This
means our two sightlines go through different absorbers or different
regions of a single absorber. In either event, we cannot conclude that
these are intrinsic NALs based only on this analysis.  On the other
hand, it is intriguing that no NAL pairs that are classified into
class-A/B have a common ejection velocity or line profile.  For
example, the \ion{C}{4} NALs at \zabs\ = 1.7652 toward image~A and at
\zabs\ = 1.7650 toward image~B are both classified as class-A NALs.
Although their ejection velocities are very close each other, their
line centers and profiles are obviously different, as shown in
Figure~\ref{f7}.  We also note that the difference is remarkable in
all three epochs (see Figure~\ref{f8}).  This means the
multi-sightline scenario (i.e., that two sightlines pass through
different regions of the outflow) is also applicable for intrinsic
NALs with large ejection velocity as well as for PALs as already
confirmed in \citet{mis13,mis14b}.

\subsubsection{Time Variability of NALs}
Because of the lower data quality (S/N $\sim$ 10~pixel$^{-1}$) and
narrower effective wavelength coverage of the Subaru/HDS spectra taken
in epochs E1 and E3, we cannot monitor NALs for time variability
analysis with only a few exceptions.  As an example, we compare
spectra obtained at three epochs around the strong \ion{C}{4} NALs at
\zabs\ = 1.8909--1.9138 (ID = 35--44) in image~A and \ion{C}{4} NALs
at \zabs\ = 1.8909--1.9119 (ID = 41--48) in image~B as shown in
Figure~\ref{f9}.  These NALs are obviously not variable, which is
consistent with the past result that NALs with large ejection
velocities are rarely variable \citep{che15}.

\subsection{Proximity Absorption Lines (PALs)}
The absorption line group at \zabs\ $\sim$ \zem\ with an ejection
velocity of \vej\ $<$ 1000~\kms\ has already been observed twice in
2010 February (epoch E1) and 2014 April (epoch E3) with Subaru/HDS.
Its origin is probably in the outflowing gas because: i) there is
evidence for partial coverage (e.g., there exists a clear residual
flux at the bottom of the \lya\ and \ion{N}{5} absorption lines even
though they appear to be saturated.), ii) the profiles are variable,
and iii) the profiles show signatures of line-locking
\citep{mis13,mis14b}.  The most important result from these past
observations is that the absorption profiles of the \lya, \ion{N}{5},
and \ion{C}{4} PALs in the spectra of images~A and B are clearly
different, especially for lines at \vej\ $<$ 0~\kms\ (see
Figure~\ref{f10}).  There are at least three possible origins for the
difference: a) a micro-lensing effect, b) time delay between the
images, and c) different absorber structure along the different
sightlines.  Among these, the first idea is immediately rejected
because the lensed images show a common ratios between the radio,
optical, and X-ray fluxes \citep{ota12,ogu12}, which is not expected
for micro-lensing.  We can also reject the time delay effect because
the variability is almost negligible between epochs E1 and E3
\citep{mis14b}.  A difference in column density between the two
sightlines is the only acceptable explanation.  To our knowledge, this
is the first time that an outflowing wind has been observed along
multiple sightlines.  Hereafter, we call the PALs at \vej\ $<$
0~\kms\ (showing sightline difference) narrow PALs and those at
\vej\ $>$ 0~\kms\ (showing similar line profiles between sightlines)
broad PALs (see Figure~\ref{f10}).

Our observation with VLT/UVES not only gives an additional epoch for
monitoring the PALs but enables us to study the detailed velocity
structure of PALs using higher quality spectra with a S/N ratio double
that of the past observations with Subaru.  We first confirm the
line-locking pattern in velocity plots of the \ion{C}{4} PAL as shown
in Figure~\ref{f11}, which was already noted in \citet{mis13}.  Like
the other NAL absorbers, the PAL absorbers also appear to be
outflowing almost parallel to our sightline.  We also reconfirm that
the \ion{C}{4} PALs became shallower (i.e., decreasing in equivalent
width) over a large velocity range of the profile between epochs E1
and E2/E3 (see Figure~\ref{f11}) at $\Delta t_{\rm rest}$ $\sim$1.3
years in the quasar rest-frame.  However, these lines are almost
stable on the shorter time scales of $\Delta t_{\rm rest}$ $\sim$0.05
years between epochs E2 and E3. The broad spectral coverage of the
VLT/UVES spectra allows us to detect for the first time the \ion{O}{6}
doublet corresponding to the PAL absorber (see
Figure~\ref{f10}). There is also a hint of the \ion{Si}{4} doublet
detected, but it appears blended with other lines at lower redshift.

\subsubsection{Comparison of Covering Factors}
In addition to line profiles and strengths, we also compare covering
factors for the clean part of the \ion{C}{4} and \ion{N}{5} PALs in
the three epochs.  The fitting parameters to the PALs in epoch E2 with
{\sc minfit} are summarized in Table~\ref{t3}.  The total column
densities of \ion{C}{4} and \ion{N}{5} PALs after summing all Voigt
components are $\log (N_{\rm N V}/{\rm cm^{-2}})$ = 15.62 and $\log
(N_{\rm C IV}/{\rm cm^{-2}})$ = 15.90 in image~A and 16.04 and 15.89
in image~B spectrum. These values are all consistent with the
corresponding values that we measured in epoch E1 \citep{mis13} with
differences of a factor of $\leq$ 2.  Because all PALs have small
offset velocities and are located on the BELs in the spectra, we
should consider the BELRs as well as the continuum source as the
background flux source.  Because the \ion{C}{4} PAL is partially
self-blended (i.e., blending of the blue members of one doublet with
the red member of another doublet), we can compare covering factors
effectively only for the narrow PALs at \vej\ $\sim$
$-$100--0~\kms\ (corresponding to $\Delta v$ $\sim$ 400--500~\kms\ in
Figure~\ref{f12}), for which self-blending is
negligible\footnote[18]{Negative ejection velocity for these
  components could be due to our underestimation of these values
  because the emission redshift is determined from broad UV emission
  lines, as done for this quasar by \citet{ina06}, which are
  systematically blueshifted from the systemic redshift that is
  measured by narrow, forbidden lines (see, e.g.,
  \citealt{cor90,tyt92,bro94,mar96}) by about 260~\kms.}.

We show the fitting results for the PALs in epoch E2 in
Figure~\ref{f12} and Table~\ref{t3}.  The average covering factors of
\ion{C}{4} and \ion{N}{5} at \vej\ $\sim$ $-$100--0~\kms\ (i.e.,
$\Delta v$ $\sim$ 400--500~\kms\ in Figure~\ref{f12}) are \cf\ =
0.47$\pm$0.03 and 0.58$\pm$0.05 toward image~A, and \cf\ =
0.23$\pm$0.02 and 0.39$\pm$0.08 toward image~B.  These values are all
consistent with the corresponding values in epoch E1 toward image~A
(\cf\ = 0.47$\pm$0.05 and 0.61$\pm$0.07) and image~B (\cf\ $\sim$ 0.2
and 0.35$\pm$0.05)\footnote[19]{Because no absorption component was
  used for \ion{C}{4} PAL at \vej\ $\leq$ 0~\kms\ in the epoch E1
  spectrum \citep{mis13}, we adopt an average line depth for this
  region as a covering factor.} and in epoch E3 toward image~A (\cf\ =
0.45$\pm$0.09 and 0.54$\pm$0.09) and image~B (\cf\ = 0.20$\pm$0.05 and
0.34$\pm$0.19).  Thus, covering factors are not variable at least on a
timescale of $\Delta t_{\rm rest}$ $\sim$ 1.3 years.

The \ion{N}{5} PALs all have larger \cf\ values than that for
\ion{C}{4}, which is consistent with past results that higher
ionization transitions tend to have larger coverage fractions
\citep[e.g.,][]{pet99,sri00,mis07a,muz15}.  This result suggests that
i) the size (i.e., distance from the central black hole) of the
\ion{N}{5} BELR is smaller than that of the \ion{C}{4} BELR, and/or
ii) the size of the \ion{N}{5} absorbers is larger than that of the
\ion{C}{4} absorbers.  It is also interesting that the PALs in the
image~A spectrum always have larger \cf\ values than those in the
image~B spectrum, which is additional evidence for variations in the
structure along multiple sightlines.

\section{DISCUSSION}

\subsection{Velocity anisotropies in NAL absorbers}
Line-locking is seen in both NALs and PALs, thus we are likely to be
observing the quasar almost along the direction of the outflow.
Nonetheless, we discovered a velocity difference between the two
sightlines (see Table~2).  Given the redshifts of the lens (\zl\ =
0.58; \citealt{ogu08}) and the source (\zs\ = 2.197) the separation
angle of the light rays that form images~A and B, as seen from the
source\footnote[20]{The separation angle as seen by the source is
  given by $\theta^{\prime}$ = $(D_{\rm ol}/D_{\rm sl}) \times
  \theta$, where $\theta$ is the observed separation angle of the
  images (22.$\!\!^{\prime\prime}$5) and $D_{\rm ol}$ and $D_{\rm sl}$
  = $((1+z_s)/(1+z_l)) \times D_{\rm ls}$ are angular diameter
  distances from the observer to the lens and from the source to the
  lens, respectively.}, is $\theta^{\prime}$
$\sim$14.$\!\!^{\prime\prime}$6.  This velocity gradient suggests that
the outflowing wind has internal velocity anisotropies on a scale of
$\sim$1~\kms~arcsec$^{-1}$ on average.  Indeed, the hydrodynamic
simulations of \citet{pro04} show this type of internal velocity
variations.  A recent radiation-MHD simulation by \citet{tak13} also
reproduced such velocity variations over a typical spatial scale of
$\sim$ 5$\times$10$^{-4}$~pc.  This is consistent with the size of NAL
absorbers, which is also comparable to the size of the continuum
source ($\sim$ 2.5$\times$10$^{-4}$~pc; \citealt{mis13}).  However, we
should note that the velocity variations in the simulations are found
in the inner part of the wind at distances of $\sim$50$R_{\rm g}$ from
the center.

The velocity distribution of intrinsic NALs that are classified into
Class-A or B appears to cluster around values of \vej\ $\sim$ 59,000,
43,000, and 29,000 \kms, except for two intrinsic NALs at \vej\ $\sim$
49,500 \kms\ (class-A) and $\sim$ 8,500 \kms\ (class-B) in the image~B
spectrum (Figure~\ref{f4}).  These clustering patterns are reminiscent
of the filamentary structures obtained by numerical simulations
\citep[e.g.,][]{pro00}.  If these patterns are indeed due to
filamentary structures, there should exist some velocity anisotropies
within them. Velocity dispersions in the three intrinsic NAL clusters
are $\delta v$ $\sim$ 900, 260, and 1200 \kms\ respectively, which
correspond to about 1.6\%, 0.6\%, and 4.0\%\ of their average ejection
velocity.

\subsection{Ionization condition in PAL absorbers}
The \lya, \ion{C}{4}, and \ion{N}{5} PALs have been monitored at three
epochs (E1, E2, and E3) between 2010 February and 2014 April.  Between
epochs E1 and E2/E3, \ion{C}{4} PALs show a clear variation in their
strength (i.e., depth).  There are several possible reasons for this
time variability: (a) gas motion across our line of sight
\citep[e.g.,][]{ham08,gib08,muz15}, (b) changes in the ionization
state of the absorber \citep[e.g.,][]{ham11,mis07b}, and (c)
redirection of photons around the absorber by scattering material
\citep[e.g.,][]{lam04,mis10}. Among these, the first scenario can be
rejected because all absorption components in the \ion{C}{4} PAL vary
in concert, which requires the implausible situation in which all
clouds cross our sightline simultaneously \citep[c.f.][]{mis07b}. The
third scenario is also less likely because it requires a variation in
the covering factor while the \cf\ values remain almost stable both in
\ion{C}{4} and \ion{N}{5} PALs between epochs E1 and E2/E3.  Thus,
only the scenario involving a change in ionization state deserves
further investigation.  This scenario is further separated into two
variants: i.e., C$^{3+}$ ions are ionized to C$^{4+}$ or they
recombine to C$^{2+}$.  Both variants of this scenario can explain the
decreasing EW of the \ion{C}{4} PALs.  Without knowing an absorber's
ionization parameter\footnote[21]{The ionization parameter $U$ is
  defined as the ratio of hydrogen ionizing photon density
  ($n_{\gamma}$) to the electron density ($n_e$).}, we cannot tell
which one is more likely.  If the latter variant applies, we can place
constraints on the electron density and the distance from the ionizing
photon source by the same prescription as used in \citet{nar04},
taking the variability time scale as an upper limit to the
recombination time.  Based on the observation that the \ion{C}{4} PALs
vary between epochs E1 and E3 over $\Delta t_{\rm obs}$ = 1514~days
(i.e., $\Delta t_{\rm rest}$ = 474~days)\footnote[22]{We compare
  spectra in epochs E1 and E3 instead of epochs E1 and E2 because the
  E2 observation spans the time range 2014 January 28 to February 26,
  which gives an additional uncertainty for measuring the variation
  time scale.} we can place a lower limit on the electron density of
the absorber as $n_e > 8.7\times 10^{3}$~\cmmm, and an upper limit on
the distance from the flux source as $r < 620$~pc, assuming $U \sim
0.02$, the value at which the \ion{C}{4} and \ion{N}{5} ions are close
to the optimal ionization states for those elements
\citep[e.g.,][]{ham95}.

\subsection{Global Picture of the Outflow from SDSS~J1029+2623}
In Table~\ref{t5}, we summarize the physical properties of broad PALs,
narrow PALs, and intrinsic NALs.  We also present a possible geometry
of the outflow along our two sightlines in Figure~\ref{f13} based on
our previous constraints.

First, we see almost the same absorption profiles of \lya, \ion{N}{5},
and \ion{C}{4} PALs in the two lensed images except for a clear
difference in the narrow PALs at \vej\ $\sim$ $-$100--0~\kms.  We also
confirm that none of the intrinsic NALs have common absorption
profiles between the images. These results suggest that the size of
the broad PAL absorbers are larger than the projected separation
between sightlines at the distance of the absorbers, $r\theta$, while
the narrow PAL absorbers and NAL absorbers have sizes smaller than
$r\theta$.  Such common absorption profiles are, however, observable
regardless of the absorber's size, if their distance from the flux
source is smaller than the boundary radius $r_{\rm b}$ at which the
two sightlines of lensed images become fully separated with no overlap
\citep{mis13}.  The boundary distance is $\sim$3.5~pc for
SDSS~J1029+2623 if only the continuum source is counted.  If we also
consider the BELR (whose size is estimated to be
$\sim$0.09~pc)\footnote[23]{This is calculated by \citet{mis13} based
  on the empirical equation in \citet{mcl04}.} as the background
source, $r_{\rm b}$ would be $\sim$1200~pc (see Figure~\ref{f13}).

We can also place constraints on the absorber size based on partial
covering analysis.  Intrinsic NALs with large ejection velocities have
partial coverage, although they absorb only the continuum
photons. This means their physical scale is comparable to or smaller
than the size of the continuum source, $d_{\rm cloud}$ $\leq$
2.5$\times$10$^{-4}$~pc. On the other hand, broad PAL absorbers cover
almost entirely both the BELR and the continuum source, which means
the size of the absorbers as a whole is comparable to or larger than
the BELR size, $\geq 0.09$~pc. Narrow PAL absorbers may consist of a
number of small clumpy clouds because they show partial coverage.

In our VLT/UVES spectra, we detected high-ionization transitions like
\ion{O}{6} and \ion{N}{5} in the PAL systems, while the intrinsic NAL
systems are in various ionization states, with or without
high-ionization transitions, as already noted in the literature
\citep[e.g.,][]{mis07a,gan13}.  Some \ion{C}{4} NAL absorbers also
show \ion{N}{5} transitions, while others do not.  Because \lya\ and
\ion{N}{5} in the PAL system are less variable than \ion{C}{4} and
because \ion{N}{5} has larger covering factor than \ion{C}{4}, the
cross-section of \lya\ and \ion{N}{5} should be larger than that of
\ion{C}{4}.  PAL absorbers with strong \ion{O}{6} lines are probably
located much closer to the ionizing flux source than NAL absorbers,
although this conclusion depends on the density of the absorber.

It is well known that both broad and narrow absorption lines at
\zabs\ $\sim$ \zem\ tend to vary \citep{wis04,mis14a}, while NALs with
large ejection velocities are rarely variable
\citep[e.g.,][]{che15}. Indeed, the PALs in our spectra are variable,
while most of the \ion{C}{4} NALs are probably stable between the
three epochs, as we observed for a few strong \ion{C}{4} NALs.
\citet{mis14a} suggest that broader absorption lines like the broad
PALs can vary mainly due to a change in their ionization state while
narrow absorption lines like narrow PALs and NALs vary primarily due
to the gas motion transverse to our sightlines.  The gas motion
scenario could explain why narrow PALs are variable but high-velocity
NALs are not.  Based on the dynamical model of \citet{mur95} and more
recent investigations \citep[e.g.,][]{mis05,hal11}, the absorbers at
larger distance from the center have small transverse (i.e., orbital)
velocities compared to their radial velocities.  If intrinsic NALs lie
at larger distances than narrow PALs, their small transverse
velocities would rarely lead to time variability, while the large
transverse velocities of narrow PALs can lead to time variability more
frequently.

\section{SUMMARY}

In this study, we performed a spectroscopic observation for images~A
and B of the gravitationally lensed quasar SDSS~J1029+2623, and
monitored the absorption profiles in these spectra as well as in the
previous two observations.  Using high quality spectra taken with
VLT/UVES, we detected intrinsic narrow absorption lines (NALs) as well
as broader, proximity absorption lines (PALs), and fit models to the
line profiles.  Based on the results of our multi-sightline
spectroscopy, we discuss a possible geometry and internal structure of
the outflowing wind along our sightlines.  Our main results are as
follows.

\begin{itemize}

\item We detected 66 NALs, of which 24 are classified as {\it
  intrinsic} NALs (physically associated with the quasar) based on
  partial coverage analysis.

\item Class-A and B NALs cluster at \vej\ $\sim$ 59,000, 43,000, and
  29,000 \kms, which is reminiscent of the filamentary structures that
  are often obtained in numerical simulations.

\item Our multiple sightline observation suggests that the size of the
  broad PAL absorbers are larger than the projected distance between
  sightlines ($r\theta$) while the narrow PAL absorbers and intrinsic
  NAL absorbers have sizes smaller than $r\theta$ if their radial
  distances are greater than the boundary distance.

\item While PAL systems show only high ionization transitions,
  including \ion{O}{6}, intrinsic NAL systems show a wide range of
  ionization conditions with and without low ionization transitions
  like \ion{O}{1}, \ion{Al}{2}, and \ion{Si}{2}, as noted in the
  literature.

\item No class-A/B NALs (i.e., candidates for intrinsic NALs) have
  common absorption profiles in the two lensed images, which means
  that the outflow has an internal velocity structure whose typical
  spatial scale is smaller than the physical distance between the
  sightlines (i.e., $\leq r\theta$).

\item Short-time variation in the PALs is probably due to a change in
  the ionization state of the gas. If this is the case, we can place a
  lower limit on the gas density as $n_e \geq 8.7 \times
  10^3$~\cmmm\ and an upper limit on the absorber's distance from the
  flux source as $r \leq\ 620$~pc.

\item Based on our best knowledge, we present a possible geometry of
  the outflow along our two sightlines, in which we identify different
  structures in the outflowing wind that can produce, respectively,
  broad PALs, narrow PALs, and NALs with large ejection velocities.

\end{itemize}

For further investigations of the outflow's internal structure,
especially in the transverse direction, we should perform the same
observations for image~C of SDSS~J1029+2623. We also aim to observe
several lensed quasars with smaller separation angles ($\theta$ $\sim$
2$^{\prime\prime}$) by a single massive galaxy to examine $\sim$10
times finer internal structure in an outflow, as already mentioned in
\citet{mis14b}.

\acknowledgments We thank the anonymous referee for comments that
helped us improve the paper.  We would like to thank Masamune Oguri,
Poshak Gandhi, and Chris Culliton for their valuable comments.  We
also would like to thank Christopher Churchill for providing us with
the {\sc minfit} and {\sc search} software packages.  The research was
supported by the Japan Society for the Promotion of Science through
Grant-in-Aid for Scientific Research 15K05020, JGC-S Scholarship
Foundation, and partially supported by MEXT Grant-in-Aid for
Scientific Research on Innovative Areas (No. 15H05894). CS
acknowledges support from CONICYT-Chile through Becas Chile 74140006.
FEB acknowledges support from CONICYT-Chile (Basal-CATA PFB-06/2007,
FONDECYT Regular 1141218, "EMBIGGEN" Anillo ACT1101) and the Ministry
of Economy, Development, and Tourism's Millennium Science Initiative
through grant IC120009, awarded to The Millennium Institute of
Astrophysics, MAS. JCC and ME acknowledge support from the National
Science Foundation through award AST-1312686.

\clearpage
\begin{deluxetable*}{ccccccc}
\tablecaption{Log of Observations \label{t1}}
\tablewidth{0pt}
\tablehead{
\colhead{Target}         &
\colhead{Obs Date}       &
\colhead{Instrument}     &
\colhead{$R$}            &
\colhead{$T_{\rm exp}$}    &
\colhead{S/N$^a$}        &
\colhead{Reference$^b$}  \\
\colhead{}               &
\colhead{}               &
\colhead{}               &
\colhead{}               &
\colhead{(sec)}          & 
\colhead{(pix$^{-1}$)}    &
\colhead{}               
}
\startdata
SDSS~J1029+2623~A & 2010 Feb 10          & Subaru/HDS & 30000 & 14400 & 13 & 1 \\
                  & 2014 Jan 28 -- Feb 3 & VLT/UVES   & 33000 & 26670 & 23 & 2 \\
                  & 2014 Apr  4          & Subaru/HDS & 36000 & 11400 & 14 & 3 \\
\hline
SDSS~J1029+2623~B & 2010 Feb 10          & Subaru/HDS & 30000 & 14200 & 13 & 1 \\
                  & 2014 Feb  4 -- 26    & VLT/UVES   & 33000 & 26670 & 23 & 2 \\
                  & 2014 Apr  4          & Subaru/HDS & 36000 & 11400 & 14 & 3 \\
\enddata
\tablenotetext{a}{Signal to noise ratio at $\lambda_{\rm obs}$ $\sim$
  4700\AA.}
\tablenotetext{b}{References. (1) \citealt{mis13}, (2) This paper, (3)
  \citealt{mis14b}.}
\end{deluxetable*}

\clearpage
\begin{landscape}
\LongTables
\begin{deluxetable*}{cccccccccccccccccr}
\tablecaption{Absorption Systems with Doublet Lines \label{t2}}
\tablewidth{0pt}
\tablehead{
\multicolumn{8}{c}{Image~A}            &
\multicolumn{1}{c}{}                   &
\multicolumn{8}{c}{Image~B}            &
\multicolumn{1}{c}{}                   \\
\cline{1-8} 
\cline{10-17}
\colhead{Ion}                          &
\colhead{\zabs}                        &
\colhead{\vej}                         &
\colhead{EW$_{\rm b}^{\rm rest}$$^a$}      &
\colhead{EW$_{\rm r}^{\rm rest}$$^b$}      &
\colhead{ID}                           &
\colhead{Class-1$^c$}                  &
\colhead{Class-2$^d$}                  &
\colhead{}                             &
\colhead{Ion}                          &
\colhead{\zabs}                        &
\colhead{\vej}                         &
\colhead{EW$_{\rm b}^{\rm rest}$$^a$}      &
\colhead{EW$_{\rm r}^{\rm rest}$$^b$}      &
\colhead{ID}                           &
\colhead{Class-1$^c$}                  &
\colhead{Class-2$^d$}                  &
\colhead{$\Delta v$$^e$}               \\
\colhead{}                             &
\colhead{}                             &
\colhead{(\kms)}                       &
\colhead{(\AA)}                        &
\colhead{(\AA)}                        &
\colhead{}                             &
\colhead{}                             &
\colhead{}                             &
\colhead{}                             &
\colhead{}                             &
\colhead{}                             &
\colhead{(\kms)}                       &
\colhead{(\AA)}                        &
\colhead{(\AA)}                        &
\colhead{}                             &
\colhead{}                             &
\colhead{}                             &
\colhead{(\kms)}                       
}
\startdata
\ion{Mg}{2}  &  0.5111 & 190431 & 0.604$\pm$0.008 & 0.495$\pm$0.008 & 27,29 & C1   & L    & &             &        &        &                 &                 &       &      &      &        \\ 
\hline                                                                                                                                                          
             &         &        &                 &                 &       &      &      & & \ion{Mg}{2} & 0.5124 & 190276 & 0.095$\pm$0.004 & 0.052$\pm$0.004 & 31,32 & C3   & L    &    0.0 \\ 
\ion{Mg}{2}  &  0.5125 & 190265 & 0.221$\pm$0.005 & 0.145$\pm$0.004 & 28,30 & C1   & L    & &             &        &        &                 &                 &       &      &      &  +19.8 \\ 
\hline                                                                                                                                                          
             &         &        &                 &                 &       &      &      & & \ion{Mg}{2} & 0.6731 & 171003 & 1.689$\pm$0.008 & 1.444$\pm$0.008 & 49,50 & C1   & L    &        \\ 
\hline                                                                                                                                                          
\ion{Mg}{2}  &  0.9176 & 141245 & 0.193$\pm$0.002 & 0.120$\pm$0.002 & 61,63 & C3   & L    & &             &        &        &                 &                 &       &      &      &    0.0 \\ 
             &         &        &                 &                 &       &      &      & & \ion{Mg}{2} & 0.9184 & 141157 & 0.186$\pm$0.005 & 0.092$\pm$0.005 & 67,68 & C1   & L    & +125.1 \\ 
\ion{Mg}{2}  &  0.9187 & 141116 & 1.472$\pm$0.006 & 1.366$\pm$0.006 & 62,64 & C1   &      & &             &        &        &                 &                 &       &      &      & +172.0 \\ 
\hline                                                                                                                                                          
             &         &        &                 &                 &       &      &      & & \ion{C}{4}  & 1.6085 &  60206 & 0.035$\pm$0.002 & 0.026$\pm$0.002 &  7,10 & B2   & H    &        \\ 
\hline                                                                                                                                                          
             &         &        &                 &                 &       &      &      & & \ion{C}{4}  & 1.6149 &  59493 & 0.105$\pm$0.004 &     $\leq$0.132 & 12,13 & B2   & IH   &    0.0 \\ 
\ion{C}{4}   &  1.6151 &  59472 & 0.308$\pm$0.003 & 0.297$\pm$0.003 &  8,11 & B2,H & IH   & &             &        &        &                 &                 &       &      &      &  +22.9 \\ 
\ion{C}{4}   &  1.6160 &  59378 & 0.047$\pm$0.003 &     $\leq$0.033 &  9,12 & C2   &      & &             &        &        &                 &                 &       &      &      & +126.2 \\ 
\hline                                                                                                                                                          
             &         &        &                 &                 &       &      &      & & \ion{C}{4}  & 1.6221 &  58699 &     $\leq$0.284 & 0.144$\pm$0.002 & 16,20 & C2,H & H    &    0.0 \\ 
\ion{C}{4}   &  1.6230 &  58601 &     $\leq$0.512 & 0.190$\pm$0.007 & 16,18 & B2,H & H    & &             &        &        &                 &                 &       &      &      & +103.0 \\ 
\hline                                                                                                                                                          
             &         &        &                 &                 &       &      &      & & \ion{C}{4}  & 1.6253 &  58350 & 0.021$\pm$0.002 & 0.014$\pm$0.002 & 18,22 & B2   & H    &        \\ 
\hline                                                                                                                                                          
             &         &        &                 &                 &       &      &      & & \ion{C}{4}  & 1.6556 &  55033 & 0.133$\pm$0.004 & 0.081$\pm$0.003 & 25,26 & C1   & H    &        \\ 
\hline                                                                                                                                                          
             &         &        &                 &                 &       &      &      & & \ion{C}{4}  & 1.6924 &  51029 & 0.084$\pm$0.002 & 0.046$\pm$0.002 & 27,28 & C2   & H    &    0.0 \\ 
\ion{C}{4}   &  1.6930 &  50965 & 0.498$\pm$0.005 & 0.185$\pm$0.007 & 23,24 & C2,H & LIH  & &             &        &        &                 &                 &       &      &      &  +66.8 \\ 
\hline                                                                                                                                                          
             &         &        &                 &                 &       &      &      & & \ion{C}{4}  & 1.7065 &  49509 & 0.042$\pm$0.002 & 0.032$\pm$0.002 & 29,30 & A2   & H    &        \\ 
\hline                                                                                                                                                          
\ion{C}{4}   &  1.7212 &  47932 & 0.493$\pm$0.005 & 0.053$\pm$0.004 & 25,26 & C2,H & H    & &             &        &        &                 &                 &       &      &      &        \\ 
\hline                                                                                                                                                          
             &         &        &                 &                 &       &      &      & & \ion{C}{4}  & 1.7617 &  43596 & 0.010$\pm$0.002 & 0.011$\pm$0.002 & 33,36 & C3   & LIH  &    0.0 \\ 
             &         &        &                 &                 &       &      &      & & \ion{C}{4}  & 1.7627 &  43496 & 0.094$\pm$0.002 & 0.066$\pm$0.002 & 34,37 & B1   &      & +108.6 \\ 
\hline                                                                                                                                                          
             &         &        &                 &                 &       &      &      & & \ion{C}{4}  & 1.7650 &  43250 & 0.041$\pm$0.002 & 0.022$\pm$0.002 & 35,38 & A2   & H    &    0.0 \\ 
\ion{C}{4}   &  1.7652 &  43224 & 0.047$\pm$0.002 & 0.035$\pm$0.002 & 31,32 & A2   & LIH  & &             &        &        &                 &                 &       &      &      &  +21.7 \\ 
\hline                                                                                                                                                          
\ion{C}{4}   &  1.8909 &  30091 & 0.355$\pm$0.004 & 0.309$\pm$0.003 & 35,37 & C1,H & LIH  & &             &        &        &                 &                 &       &      &      &    0.0 \\ 
             &         &        &                 &                 &       &      &      & & \ion{C}{4}  & 1.8909 &  30088 & 0.456$\pm$0.006 &     $\leq$0.501 & 41,43 & C2,H & IH   &    0.0 \\ 
\ion{Si}{4}  &  1.8909 &  30088 & 0.181$\pm$0.003 &     $\leq$0.268 &  5,10 & B2,H &      & &             &        &        &                 &                 &       &      &      &    0.0 \\ 
             &         &        &                 &                 &       &      &      & & \ion{Si}{4} & 1.8909 &  30087 & 0.121$\pm$0.004 & 0.119$\pm$0.004 &  5,14 & B2   &      &    0.0 \\ 
\hline                                                                                                                                                          
             &         &        &                 &                 &       &      &      & & \ion{Si}{4} & 1.8944 &  29737 & 0.380$\pm$0.005 &     $\leq$0.383 &  6,17 & C2,H & LIH  &    0.0 \\ 
             &         &        &                 &                 &       &      &      & & \ion{C}{4}  & 1.8945 &  29723 &     $\leq$0.669 &     $\leq$0.592 & 42,44 & C2,H &      &  +10.4 \\ 
\ion{C}{4}   &  1.8949 &  29686 & 0.413$\pm$0.004 & 0.318$\pm$0.004 & 36,39 & C2,H & LIH  & &             &        &        &                 &                 &       &      &      &  +51.8 \\ 
\ion{Si}{4}  &  1.8949 &  29685 & 0.137$\pm$0.003 &     $\leq$0.464 &  6,14 & B2   &      & &             &        &        &                 &                 &       &      &      &  +51.8 \\ 
\hline                                                                                                                                                          
             &         &        &                 &                 &       &      &      & & \ion{Si}{4} & 1.8975 &  29413 & 0.032$\pm$0.002 & 0.019$\pm$0.002 &  8,19 & B2   & LIH  &    0.0 \\ 
\ion{C}{4}   &  1.8982 &  29343 & 0.193$\pm$0.004 & 0.146$\pm$0.005 & 38,40 & C1,H & IH   & &             &        &        &                 &                 &       &      &      &  +72.5 \\ 
\ion{Si}{4}  &  1.8982 &  29342 & 0.029$\pm$0.003 & 0.013$\pm$0.002 &  7,17 & C2   &      & &             &        &        &                 &                 &       &      &      &  +72.5 \\ 
\hline                                                                                                                                                          
             &         &        &                 &                 &       &      &      & & \ion{Si}{4} & 1.9016 &  28991 & 0.162$\pm$0.003 & 0.142$\pm$0.002 &  9,21 & C1,H & LIH  &    0.0 \\ 
             &         &        &                 &                 &       &      &      & & \ion{C}{4}  & 1.9019 &  28964 & 0.457$\pm$0.005 & 0.363$\pm$0.006 & 45,46 & C1,H &      &  +31.0 \\ 
             &         &        &                 &                 &       &      &      & & \ion{Si}{4} & 1.9024 &  28917 & 0.023$\pm$0.002 & 0.013$\pm$0.002 & 11,23 & B2   &      &  +82.7 \\ 
\hline                                                                                                                                                          
\ion{C}{4}   &  1.9115 &  27981 & 0.572$\pm$0.008 & 0.448$\pm$0.007 & 41,43 & C1,H & LIH  & &             &        &        &                 &                 &       &      &      &    0.0 \\ 
\ion{Si}{4}  &  1.9115 &  27987 & 0.439$\pm$0.004 & 0.318$\pm$0.004 & 13,19 & C1,H &      & &             &        &        &                 &                 &       &      &      &    0.0 \\ 
             &         &        &                 &                 &       &      &      & & \ion{Si}{4} & 1.9118 &  27949 &     $\leq$0.248 &     $\leq$0.197 & 15,24 & C2,H &      &  +30.9 \\ 
             &         &        &                 &                 &       &      &      & & \ion{C}{4}  & 1.9119 &  27946 & 0.837$\pm$0.009 & 0.679$\pm$0.009 & 47,48 & C1,H & LIH  &  +41.2 \\ 
\cline{1-7}                                                                                                                                                     
\ion{C}{4}   &  1.9138 &  27745 & 0.484$\pm$0.009 & 0.359$\pm$0.009 & 42,44 & C3,H &      & &             &        &        &                 &                 &       &      &      & +236.9 \\ 
\ion{Si}{4}  &  1.9141 &  27718 &     $\leq$0.461 & 0.205$\pm$0.005 & 15,20 & B2,H & LIH  & &             &        &        &                 &                 &       &      &      & +267.8 \\ 
\hline                                                                                                                                                          
\ion{Si}{4}  &  1.9420 &  24880 & 0.041$\pm$0.002 & 0.019$\pm$0.003 & 21,22 & C2   & LI   & &             &        &        &                 &                 &       &      &      &        \\ 
\hline                                                                                                                                                          
             &         &        &                 &                 &       &      &      & & \ion{C}{4}  & 2.1081 &   8460 & 0.209$\pm$0.004 & 0.110$\pm$0.004 & 51,52 & B1,H & IH   &    0.0 \\ 
\ion{C}{4}   &  2.1084 &   8426 & 0.130$\pm$0.002 & 0.074$\pm$0.002 & 45,46 & C2   & IH   & &             &        &        &                 &                 &       &      &      &  +29.0 \\ 
\hline                                                                                                                                                          
\ion{C}{4}   &  2.1196 &   7348 &     $\leq$0.065 & 0.024$\pm$0.001 & 47,48 & C2   & LH   & &             &        &        &                 &                 &       &      &      &    0.0 \\ 
             &         &        &                 &                 &       &      &      & & \ion{C}{4}  & 2.1209 &   7223 & 0.552$\pm$0.003 & 0.470$\pm$0.003 & 53,54 & C1,H & LIH  & +125.0 \\ 
\hline                                                                                                                                                          
\ion{C}{4}   &  2.1270 &   6643 & 0.145$\pm$0.001 & 0.114$\pm$0.001 & 49,50 & C1   & H    & &             &        &        &                 &                 &       &      &      &    0.0 \\ 
             &         &        &                 &                 &       &      &      & & \ion{C}{4}  & 2.1276 &   6585 & 0.029$\pm$0.002 & 0.014$\pm$0.002 & 55,57 & C2   & IH   &  +57.5 \\ 
             &         &        &                 &                 &       &      &      & & \ion{C}{4}  & 2.1285 &   6500 & 0.183$\pm$0.002 & 0.155$\pm$0.002 & 56,58 & C1,H &      & +143.9 \\ 
\hline                                                                                                                                                          
\ion{C}{4}   &  2.1349 &   5884 & 0.093$\pm$0.003 & 0.046$\pm$0.002 & 51,52 & C1   & H    & &             &        &        &                 &                 &       &      &      &    0.0 \\ 
             &         &        &                 &                 &       &      &      & & \ion{C}{4}  & 2.1349 &   5886 & 0.073$\pm$0.002 & 0.032$\pm$0.002 & 59,60 & C2   & H    &    0.0 \\ 
\hline                                                                                                                                                          
\ion{C}{4}   &  2.1800 &   1604 & 0.011$\pm$0.001 & 0.008$\pm$0.001 & 53,55 & C2   & IH   & &             &        &        &                 &                 &       &      &      &    0.0 \\ 
\ion{Si}{4}  &  2.1815 &   1458 & 0.115$\pm$0.006 & 0.041$\pm$0.005 & 33,34 & C1   &      & &             &        &        &                 &                 &       &      &      & +141.5 \\ 
             &         &        &                 &                 &       &      &      & & \ion{C}{4}  & 2.1818 &   1428 & 0.298$\pm$0.002 & 0.209$\pm$0.002 & 61,62 & C1,H & LIH  & +169.7 \\ 
\ion{C}{4}   &  2.1819 &   1416 & 0.425$\pm$0.004 &     $\leq$0.297 & 54,56 & C1,H &      & &             &        &        &                 &                 &       &      &      & +179.2 \\ 
             &         &        &                 &                 &       &      &      & & \ion{Si}{4} & 2.1821 &   1403 & 0.088$\pm$0.002 & 0.062$\pm$0.003 & 39,40 & C1   &      & +198.0 \\ 
\hline                                                                                                                                                          
             &         &        &                 &                 &       &      &      & & \ion{C}{4}  & 2.1897 &    686 & 0.848$\pm$0.003 &     $\leq$1.405 & 63,64 & A1,H & H    &    0.0 \\ 
\ion{C}{4}   &  2.1898 &    676 & 0.902$\pm$0.003 &     $\leq$1.253 & 57,58 & A1,H & H    & &             &        &        &                 &                 &       &      &      &   +9.4 \\ 
\ion{N}{5}   &  2.1900 &    657 & 0.945$\pm$0.004 &     $\leq$0.882 &  1, 3 & A1,H &      & &             &        &        &                 &                 &       &      &      &  +28.2 \\ 
             &         &        &                 &                 &       &      &      & & \ion{N}{5}  & 2.1900 &    658 & 0.888$\pm$0.004 &     $\leq$0.717 &  1, 3 & A1,H &      &  +28.2 \\ 
\hline                                                                                                                                                          
\ion{N}{5}   &  2.1955 &    141 &     $\leq$1.494 & 1.341$\pm$0.004 &  2, 4 & A1,H & H    & &             &        &        &                 &                 &       &      &      &    0.0 \\ 
             &         &        &                 &                 &       &      &      & & \ion{N}{5}  & 2.1955 &    141 &     $\leq$1.231 & 1.201$\pm$0.003 &  2, 4 & A1,H & H    &    0.0 \\ 
             &         &        &                 &                 &       &      &      & & \ion{C}{4}  & 2.1957 &    122 & 1.569$\pm$0.003 & 1.224$\pm$0.003 & 65,66 & A1,H &      &  +18.8 \\ 
\ion{C}{4}   &  2.1958 &    113 &     $\leq$1.590 & 1.303$\pm$0.003 & 59,60 & A1,H &      & &             &        &        &                 &                 &       &      &      &  +28.2 \\ 
\enddata
\tablenotetext{a}{Rest-frame equivalent width of blue component of doublet line.}
\tablenotetext{b}{Rest-frame equivalent width of red component of doublet line.}
\tablenotetext{c}{Reliability class of intrinsic lines based on partial coverage analysis, defined in \citet{mis07a}. If an equivalent width is large enough, we classify the doublet as {\it homogeneous} sample with a mark of ``H''.}
\tablenotetext{d}{Ionization class of absorption system with high (H; IP $>$ 60~eV), intermediate (I; IP = 35 -- 50~eV), and/or low (L; IP $<$ 25~eV) ionization transitions.}
\tablenotetext{e}{Velocity difference from the first doublet in each absorption {\it system}.}
\end{deluxetable*}

\clearpage
\begin{deluxetable*}{cccccccl}
\tablecaption{Line Parameters of Narrow Absorption Components \label{t3}}
\tablehead{
\colhead{} &
\colhead{$\lambda_{\rm obs}$$^a$} & 
\colhead{\zabs$^b$} &
\colhead{\vej$^c$} &
\colhead{$\log N$} &
\colhead{$\sigma(v)$/$b$$^d$} &
\colhead{} & 
\colhead{Other} \\
\colhead{Ion} & 
\colhead{(\AA)} &
\colhead{} &
\colhead{(\kms)} &
\colhead{(\cmm)} &
\colhead{(\kms)} &
\colhead{\cf$^e$} &
\colhead{Ions$^f$} 
}
\startdata
\multicolumn{8}{c}{Image~A} \\
\cline{1-8} \\
\ion{Mg}{2}        &4225.6 &0.5111    &190431 &               &47.3            &                        & \ion{Mg}{1}~$\lambda$2853, \ion{Fe}{2}~$\lambda$2600 (\ion{Fe}{2}~$\lambda$2383) \\
                   &       &0.5108    &190464 &12.96$\pm$0.06 &5.1$\pm$0.2     & 0.93$^{+0.07}_{-0.06}$    & \\
                   &       &0.5112    &190426 &13.51$\pm$0.05 &10.5$\pm$0.5    & 1.00$^{+0.06}_{-0.06}$    & \\
                   &       &0.5112    &190423 &13.69$\pm$0.33 &37.1$\pm$7.2    & 0.19$^{+0.10}_{-0.09}$    & \\
                   &       &0.5113    &190408 &12.60$\pm$0.02 &7.6$\pm$0.3     & 1.00                   & \\
\cline{1-8} \\
\ion{Mg}{2}        &4229.5 &0.5125    &190265 &               &17.9            &                        & \ion{Mg}{1}~$\lambda$2853, \ion{Fe}{2}~$\lambda$2600 (\ion{Fe}{2}~$\lambda$2344, \ion{Fe}{2}~$\lambda$2383) \\
                   &       &0.5125    &190269 &12.69$\pm$0.02 &6.0$\pm$0.4     & 1.00                   & \\
                   &       &0.5126    &190261 &12.70$\pm$0.27 &13.3$\pm$4.2    & 0.83$^{+0.46}_{-0.28}$    & \\
\cline{1-8} \\
\ion{Mg}{2}        &5376.1 &0.9176    &141245 &               &14.1            &                        & \ion{Mg}{1}~$\lambda$2853, \ion{Fe}{2}~$\lambda$2600 (\ion{Al}{3}~$\lambda$1863) \\
                   &       &0.9176    &141246 &12.95$\pm$0.02 &4.7$\pm$0.2     & 1.00                   & \\
\cline{1-8} \\
\ion{Mg}{2}        &5365.3 &0.9187    &141116 &               &68.1            &                        & \ion{Mg}{1}~$\lambda$2853, \ion{Fe}{2}~$\lambda$2344,2600 (\ion{Al}{3}~$\lambda$1863) \\
                   &       &0.9181    &141190 &12.50$\pm$0.02 &4.1$\pm$0.3     & 1.00                   & \\
                   &       &0.9185    &141146 &14.03$\pm$0.02 &16.9$\pm$0.2    & 1.00                   & \\
                   &       &0.9189    &141094 &14.94$\pm$0.07 &22.0$\pm$0.5    & 0.99$^{+0.03}_{-0.03}$    & \\
\enddata
\tablenotetext{a}{Wavelength of flux-weighted line center.}
\tablenotetext{b}{Redshift of flux weighted line center.}
\tablenotetext{c}{Ejection velocity from quasar emission redshift.}
\tablenotetext{d}{Flux-weighted line width ($\sigma(v)$) or Doppler
  parameter ($b$).}  
\tablenotetext{e}{Covering factor.}  
\tablenotetext{f}{Other single lines that are detected in the
  system. Lines in parenthesis are in \lya\ forest.}
\tablenotetext{g}{Two \ion{C}{4} PALs at \zabs\ $\sim$ 2.1898 and
  2.1958 are fitted together as a single absorption profile whose
  flux-weighted redshift is \zabs\ $\sim$ 2.1937.}
\tablenotetext{h}{Two \ion{N}{5} PALs at \zabs\ $\sim$ 2.1900 and
  2.1955 are fitted together as a single absorption profile whose
  flux-weighted redshift is \zabs\ $\sim$ 2.1937.}
\tablenotetext{i}{We cannot fit this \ion{Si}{4} NAL because both
  members of the doublet are blending with other lines.}
\tablenotetext{j}{Two \ion{C}{4} PALs at \zabs\ $\sim$ 2.1897 and
  2.1957 are fitted together as a single absorption profile whose
  flux-weighted redshift is \zabs\ $\sim$ 2.1937.}
\tablenotetext{k}{Two \ion{N}{5} PALs at \zabs\ $\sim$ 2.1900 and
  2.1955 are fitted together as a single absorption profile whose
  flux-weighted redshift is \zabs\ $\sim$ 2.1935.}
\tablecomments{Table~\ref{t3} is presented in its entirety in the
  complete version; an abbreviated version of the table is shown
  here for guidance as to its form and content.}
\end{deluxetable*}
\clearpage
\end{landscape}

\begin{deluxetable*}{lcccccccccccc}
\tablecaption{Statistical Properties of Poisson Systems of NALs \label{t4}}
\tablehead{
\multicolumn{2}{c}{} &
\multicolumn{3}{c}{\ion{C}{4}} &
\multicolumn{1}{c}{} & 
\multicolumn{3}{c}{\ion{N}{5}}  &
\multicolumn{1}{c}{} & 
\multicolumn{3}{c}{\ion{Si}{4}} \\
\noalign{\vskip 3pt}
\cline{3-5}
\cline{7-9}
\cline{11-13} 
\noalign{\vskip 3pt}
\multicolumn{2}{c}{} &
\colhead{AAL$^a$} &
\colhead{non-AAL$^b$} & 
\multicolumn{1}{c}{total} & 
\colhead{} & 
\colhead{AAL$^a$} &
\colhead{non-AAL$^b$} & 
\multicolumn{1}{c}{total} & 
\colhead{} & 
\colhead{AAL$^a$} &
\colhead{non-AAL$^b$} & 
\multicolumn{1}{c}{total} 
}
\startdata
Path length & $\delta z^c$    & 0.05           & 0.56            & 0.61            & & 0.05           & 0.01          & 0.07           & & 0.05            & 0.37           & 0.42           \\
            & $\delta\beta^c$ & 0.02           & 0.19            & 0.21            & & 0.02           & 0.00          & 0.02           & & 0.02            & 0.12           & 0.14           \\
\noalign{\vskip 6pt}
Image~A     & $N^d$           & 3$^{+2.9}_{-1.6}$ & 9$^{+4.1}_{-2.9}$ & 12$^{+4.6}_{-3.4}$ & & 2$^{+2.6}_{-1.3}$ & 0$^{+1.8}_{-0}$ & 2$^{+2.6}_{-1.3}$ & & 0$^{+1.8}_{-0}$  & 3$^{+2.9}_{-1.6}$ & 3$^{+2.9}_{-1.6}$ \\
            & $dN/dz$         & 57$^{+55}_{-31}$ & 16$^{+7.3}_{-5.3}$ & 20$^{+7.4}_{-5.6}$ & & 38$^{+50}_{-24}$  & 0$^{+132}_{-0}$ & 30$^{+39}_{-19}$ & & 0$^{+35}_{-0}$   & 8.1$^{+7.9}_{-4.4}$ & 7.1$^{+6.9}_{-3.9}$ \\
            & $dN/d\beta$     & 177$^{+172}_{-96}$ & 47$^{+22}_{-16}$ & 58$^{+22}_{-17}$  & & 118$^{+155}_{-76}$ & 0$^{+409}_{-0}$ & 93$^{+123}_{-60}$ & & 0$^{+108}_{-0}$ & 25$^{+24}_{-14}$   & 22$^{+21}_{-12}$    \\
\noalign{\vskip 6pt}
Image~B     & $N^d$           & 3$^{+2.9}_{-1.6}$ & 8$^{+4.0}_{-2.8}$  & 11$^{+4.4}_{-3.3}$ & & 2$^{+2.6}_{-1.3}$ & 0$^{+1.8}_{-0}$ & 2$^{+2.6}_{-1.3}$ & & 0$^{+1.8}_{-0}$  & 3$^{+2.9}_{-1.6}$ & 3$^{+2.9}_{-1.6}$ \\
            & $dN/dz$         & 57$^{+55}_{-31}$ & 14$^{+7.1}_{-4.9}$ & 18$^{+7.2}_{-5.3}$ & & 38$^{+50}_{-24}$  & 0$^{+132}_{-0}$ & 30$^{+39}_{-19}$ & & 0$^{+35}_{-0}$    & 8.1$^{+7.9}_{-4.4}$ & 7.1$^{+6.9}_{-3.9}$ \\
            & $dN/d\beta$     & 177$^{+172}_{-96}$ & 42$^{+21}_{-15}$ & 53$^{+21}_{-16}$  & & 118$^{+155}_{-76}$ & 0$^{+409}_{-0}$ & 93$^{+123}_{-60}$ & & 0$^{+108}_{-0}$  & 25$^{+24}_{-14}$ & 22$^{+21}_{-12}$     \\
\noalign{\vskip 6pt}
\hline			  			      
\noalign{\vskip 6pt}
M07a$^e$    & $dN/dz$          & 14.     &  6.6     &  6.9    & &  5.2    & 0.0    &  4.6    & &  3.2    &  3.7    &  3.7    \\
            & $dN/d\beta$      & 54.     & 24.      & 25.     & & 21.     & 0.0    & 18.     & & 13.     & 14.     & 14.     \\
\noalign{\vskip 6pt}
\enddata
\tablenotetext{a}{Associated absorption lines with \vej\ $\leq$
  5000~\kms\ from the quasar emission redshift.}
\tablenotetext{b}{Non-associated absorption lines with \vej\ $>$
  5000~\kms\ from the quasar emission redshift.}
\tablenotetext{c}{Total redshift and speed intervals considered in the
  determination of $dN/dz$ and $dN/d\beta$.}
\tablenotetext{d}{Number of Poisson systems (i.e., groups of NALs that
  lie within 200~\kms\ of each other).}
\tablenotetext{e}{$dN/dz$ and $dN/d\beta$ values from \citet{mis07a}.}
\end{deluxetable*}

\begin{deluxetable*}{ccccc}
\tablecaption{Properties of PALs and intrinsic NALs \label{t5}}
\tablehead{
\colhead{}               &
\colhead{PAL$^a$}        &
\colhead{PAL$^b$}        &
\colhead{NAL}            &
\colhead{reference$^c$}  \\
\colhead{}               &
\colhead{(broad)}        &
\colhead{(narrow)}       &
\colhead{}               &
\colhead{}             
}
\startdata
Absorber's Transverse Size$^d$  & $\geq$ r$\theta$       & $\leq$ r$\theta$       & $\leq$ r$\theta$        & 1 \\
Ionization Condition$^e$        & high                   & high                   & low -- high             & 1, 2, 3 \\
Radial Velocity$^f$ (\vej)     & small                  & small                  & large                   & 1 \\
Rotational Velocity$^g$ (\vrot) & large                  & large                  & small                   & 7, 8, 9  \\
Variability                     & frequently             & frequently             & very rare$^h$           & 4, 5, 6 \\
Origin of Variability           & ionization condition   & gas motion             & gas motion (if any)     & 5 \\
Covering Factor (\cf)           & $\sim$ 1               & $<$ 1                  & $<$ 1                   & 1 \\
Background flux source          & BELR                   & BELR                   & continuum source        & 1 \\
\enddata
\tablenotetext{a}{PALs at \vej\ $>$ 0~\kms\ with common absorption profiles in both sightlines.}
\tablenotetext{b}{PALs at \vej\ $<$ 0~\kms\ with sightline variation.}
\tablenotetext{c}{References. (1) This paper, (2) \citet{mis07a}, (3)
  \citet{gan13} (4) \citet{wis04}, (5) \citet{mis14a}, (6)
  \citet{che15}, (7) \citet{mur95}, (8) \citet{mis05}, (9)
  \citet{hal11}}
\tablenotetext{d}{This depends on whether sightline difference is
  observed (i.e., $\leq$ r$\theta$) or not (i.e., $\geq$ r$\theta$).}
\tablenotetext{e}{{L}ow, intermediate, and high ionization transitions
  are defined in Section 4.1.3.}
\tablenotetext{f}{Ejection velocity assuming \zem\ = 2.197.}
\tablenotetext{g}{Absorbers with small orbital radius tend to have
  larger rotational velocity (see Section 5.3).}
\tablenotetext{h}{Variation probability is very low for high ejection
  velocity NALs \citep{che15}, although we confirmed this trend only 
  for a few strong \ion{C}{4} NALs.}
\end{deluxetable*}


\begin{figure*}
 \begin{center}
  \includegraphics[width=13cm,angle=0]{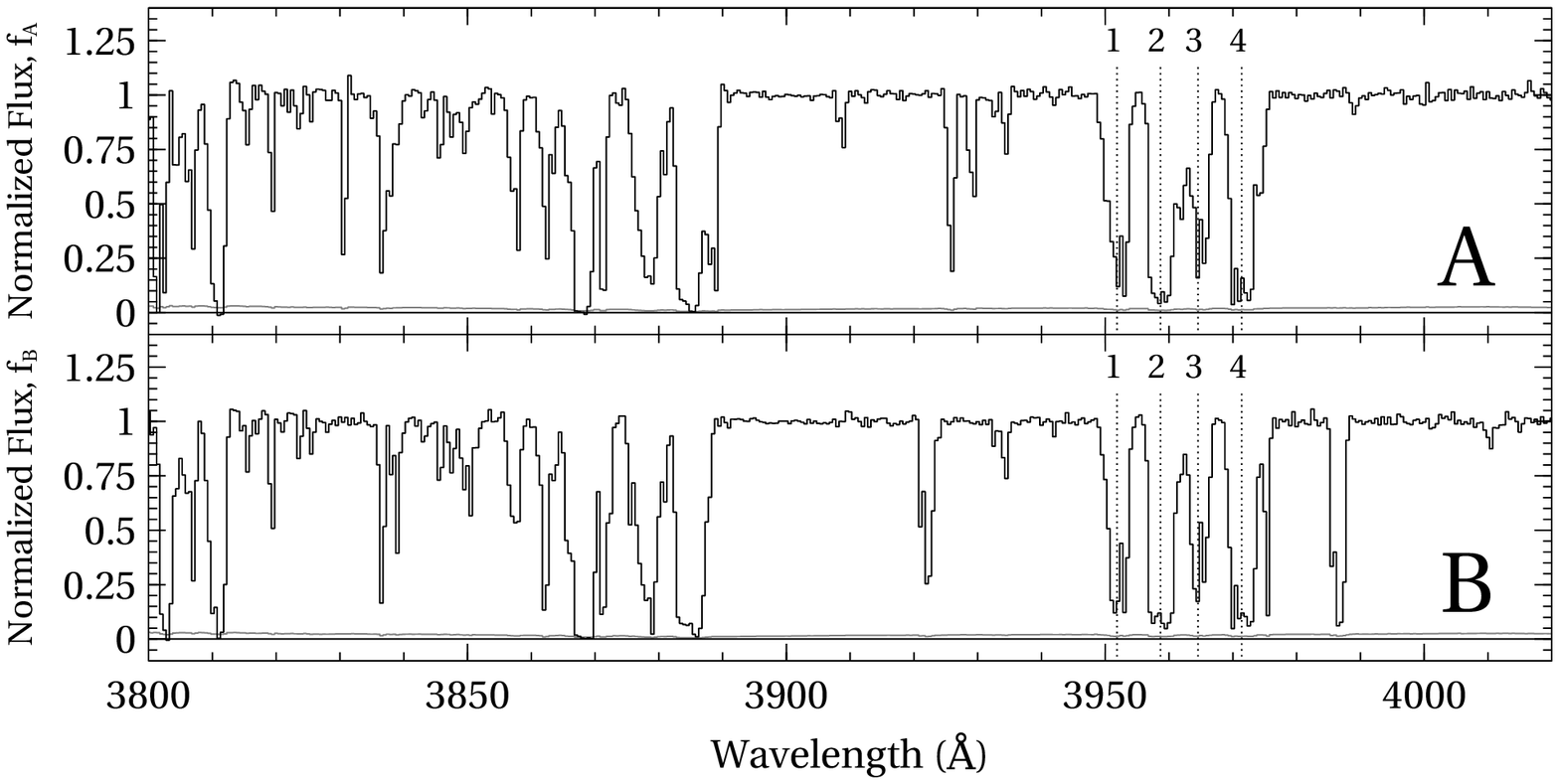}
  \includegraphics[width=13cm,angle=0]{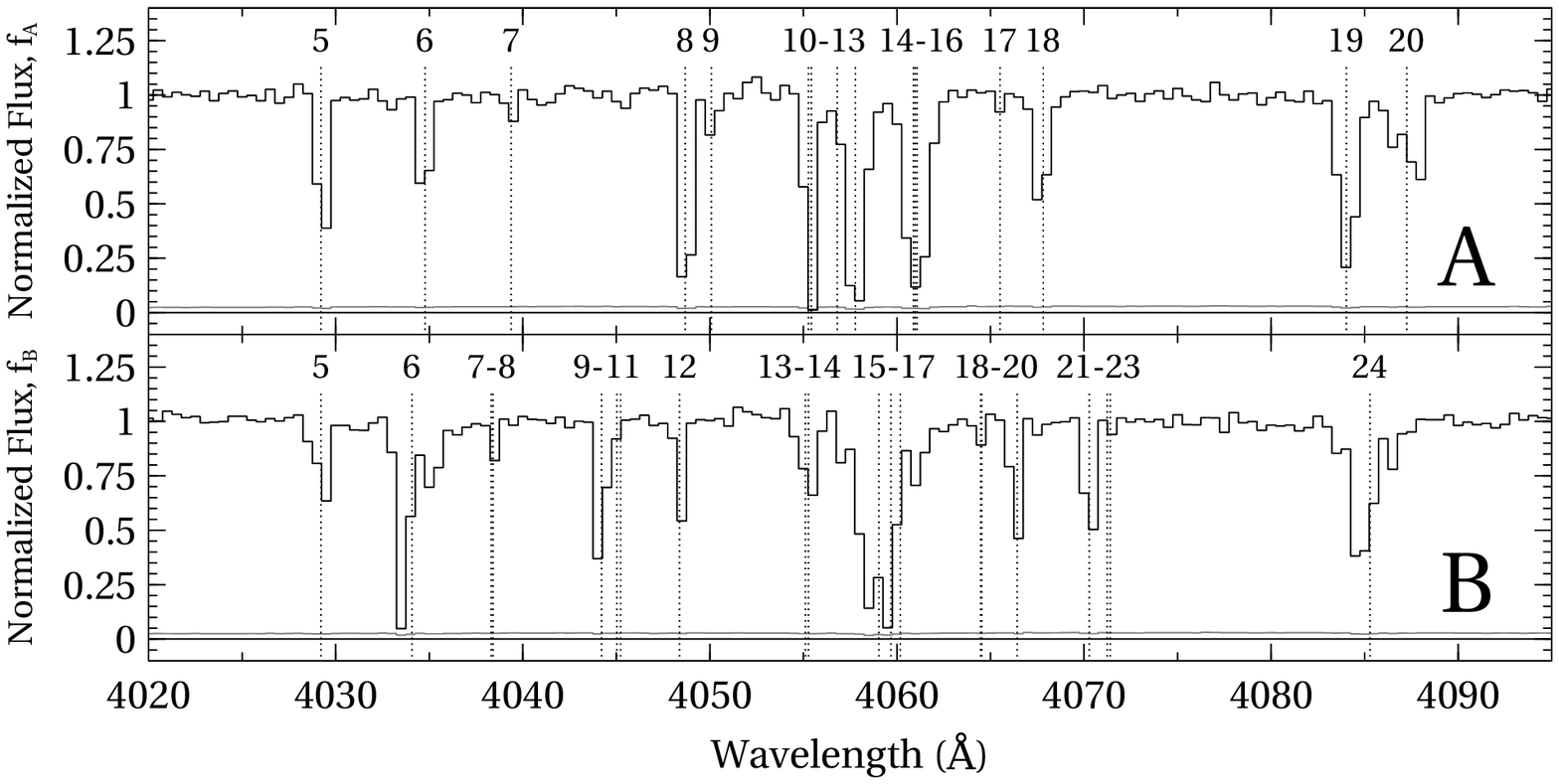}
  \includegraphics[width=13cm,angle=0]{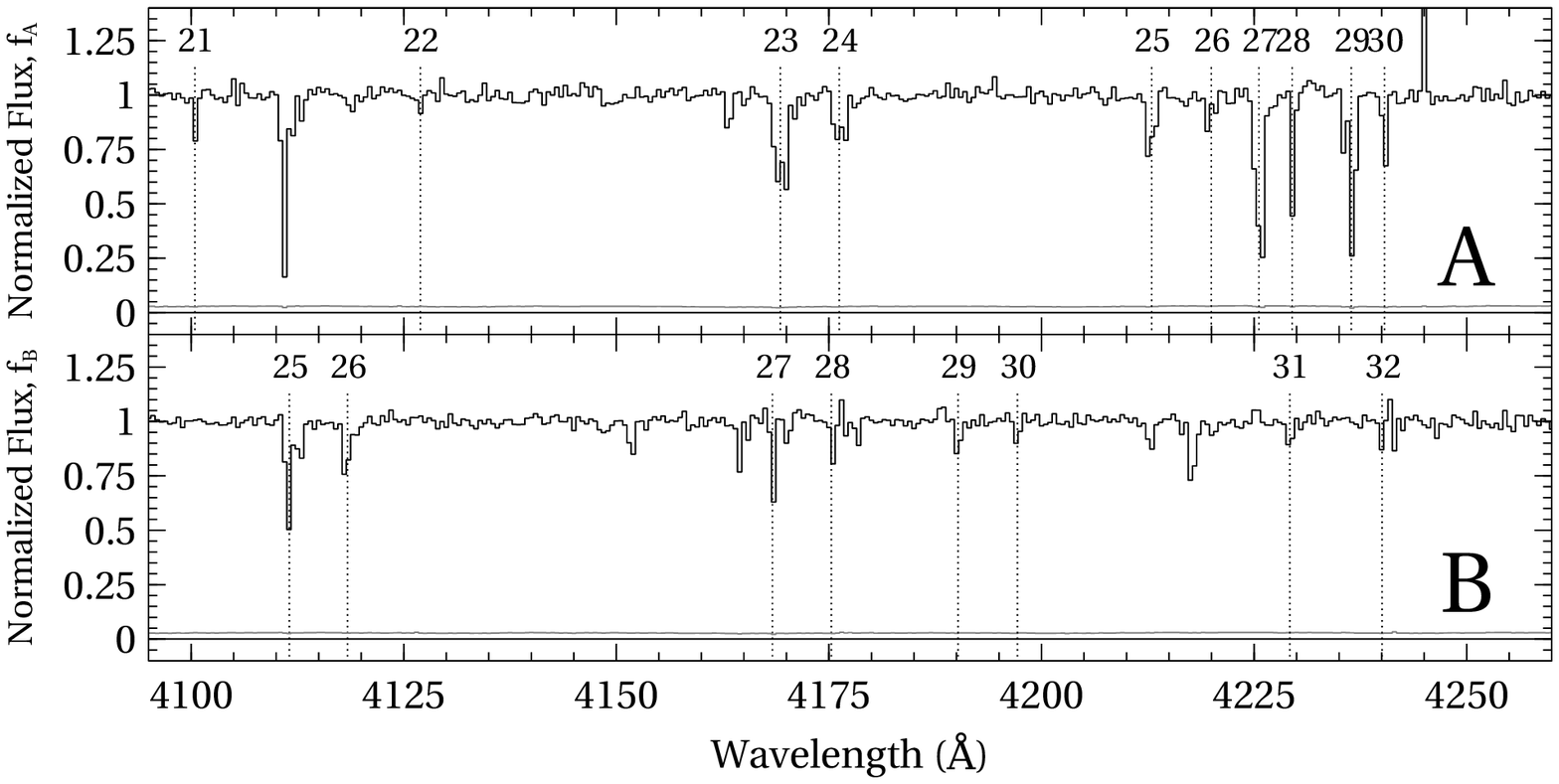}
 \end{center}
 \caption{Normalized spectra and their 1$\sigma$ errors for images~A
   and B of SDSS~J1029+2623, taken with VLT/UVES, after sampling every
   0.5\AA\ for display purpose only.\label{f1}}
\end{figure*}

\begin{figure*}
\addtocounter{figure}{-1}
 \begin{center}
  \includegraphics[width=13cm,angle=0]{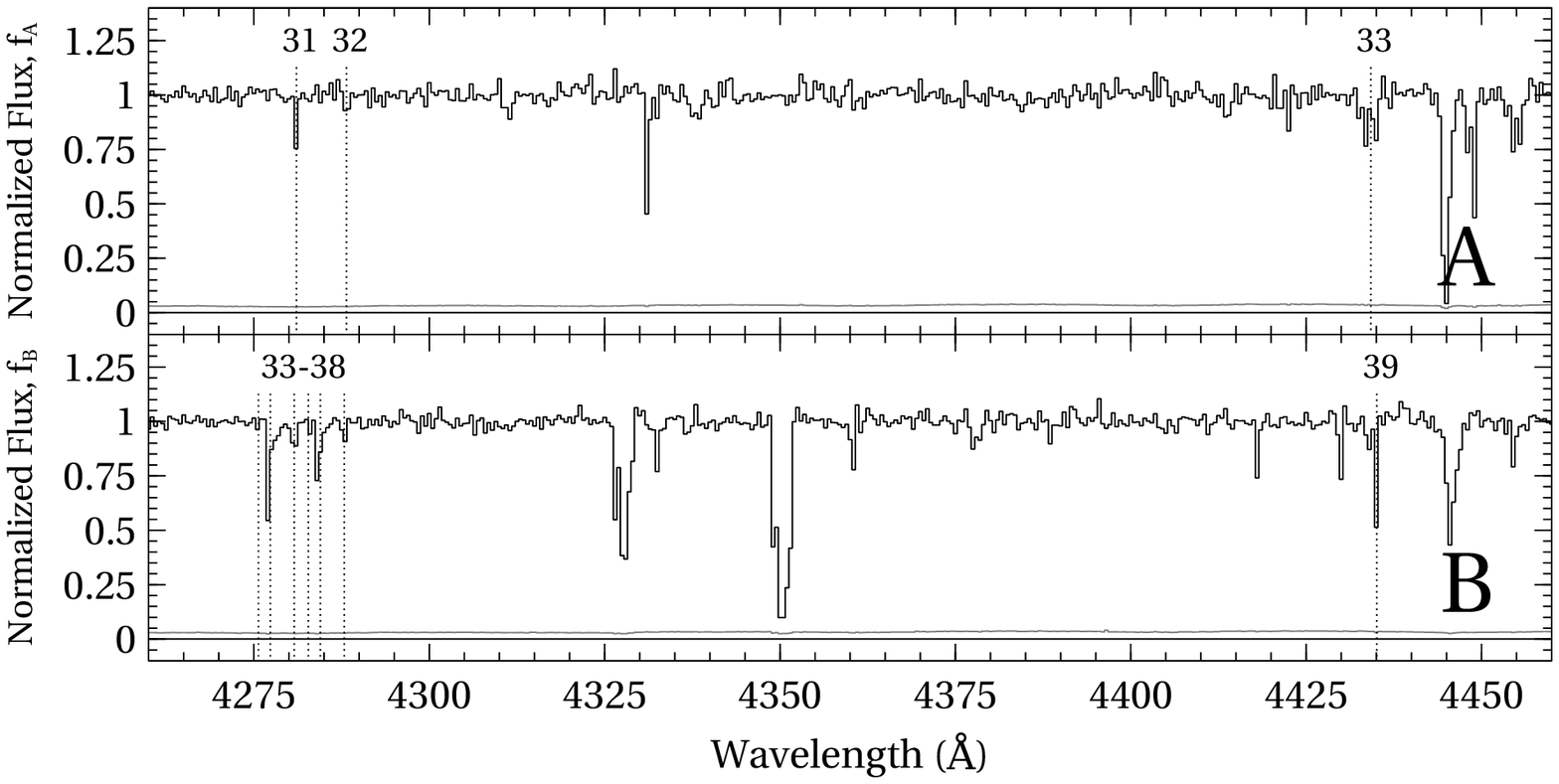}
  \includegraphics[width=13cm,angle=0]{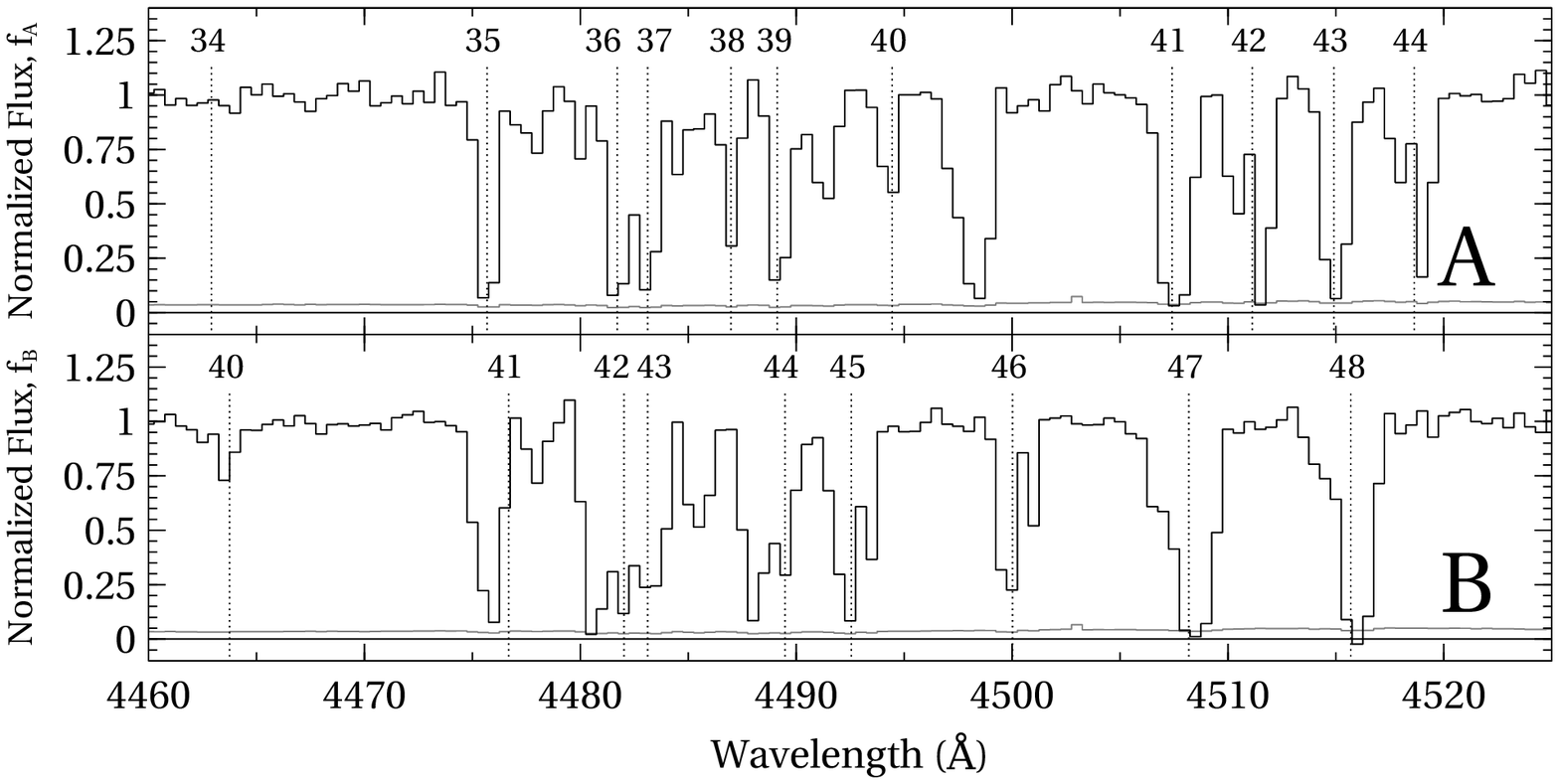}
  \includegraphics[width=13cm,angle=0]{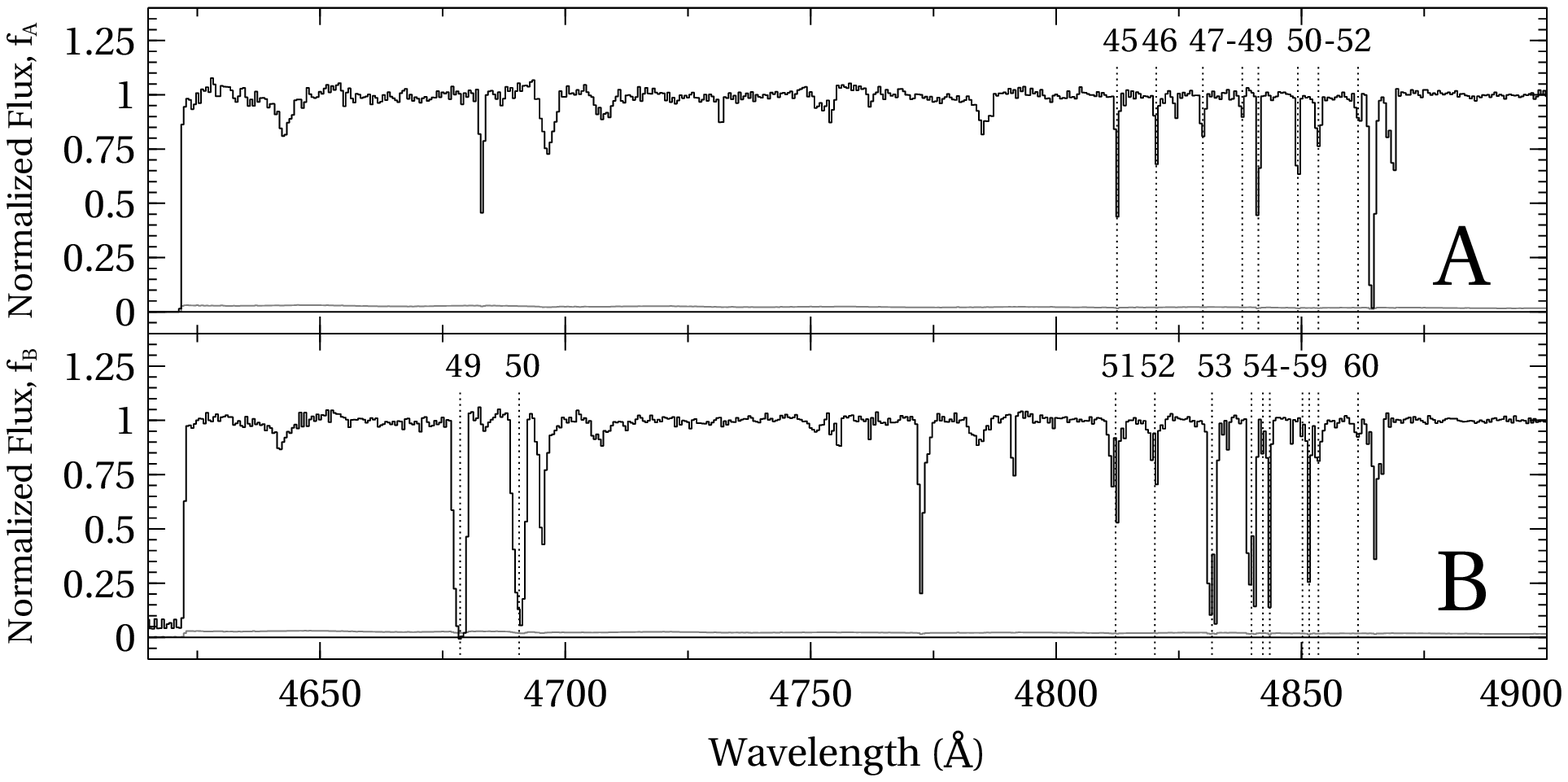}
 \end{center}
 \caption{Continued.}
\end{figure*}

\begin{figure*}
\addtocounter{figure}{-1}
 \begin{center}
  \includegraphics[width=13cm,angle=0]{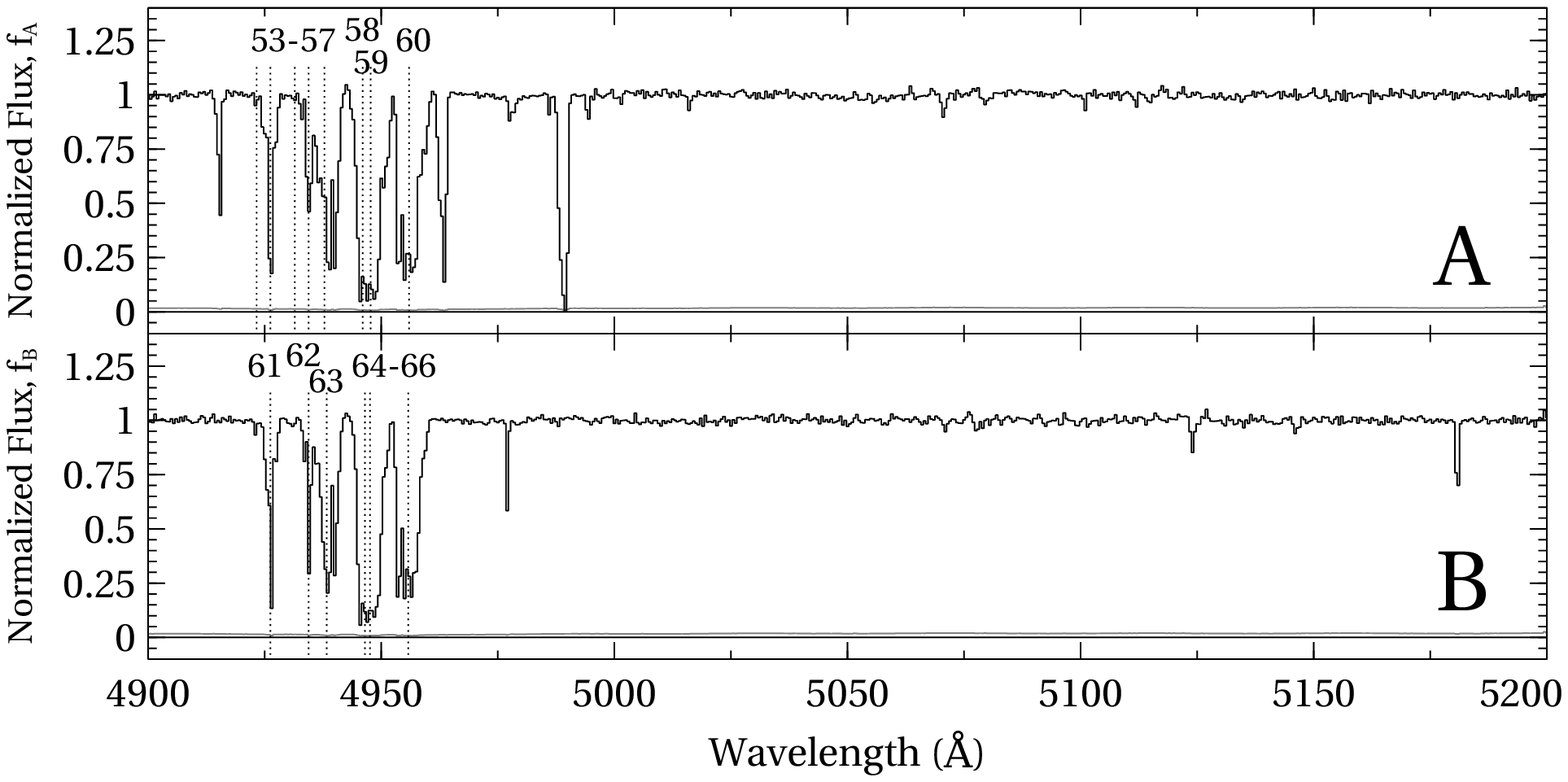}
  \includegraphics[width=13cm,angle=0]{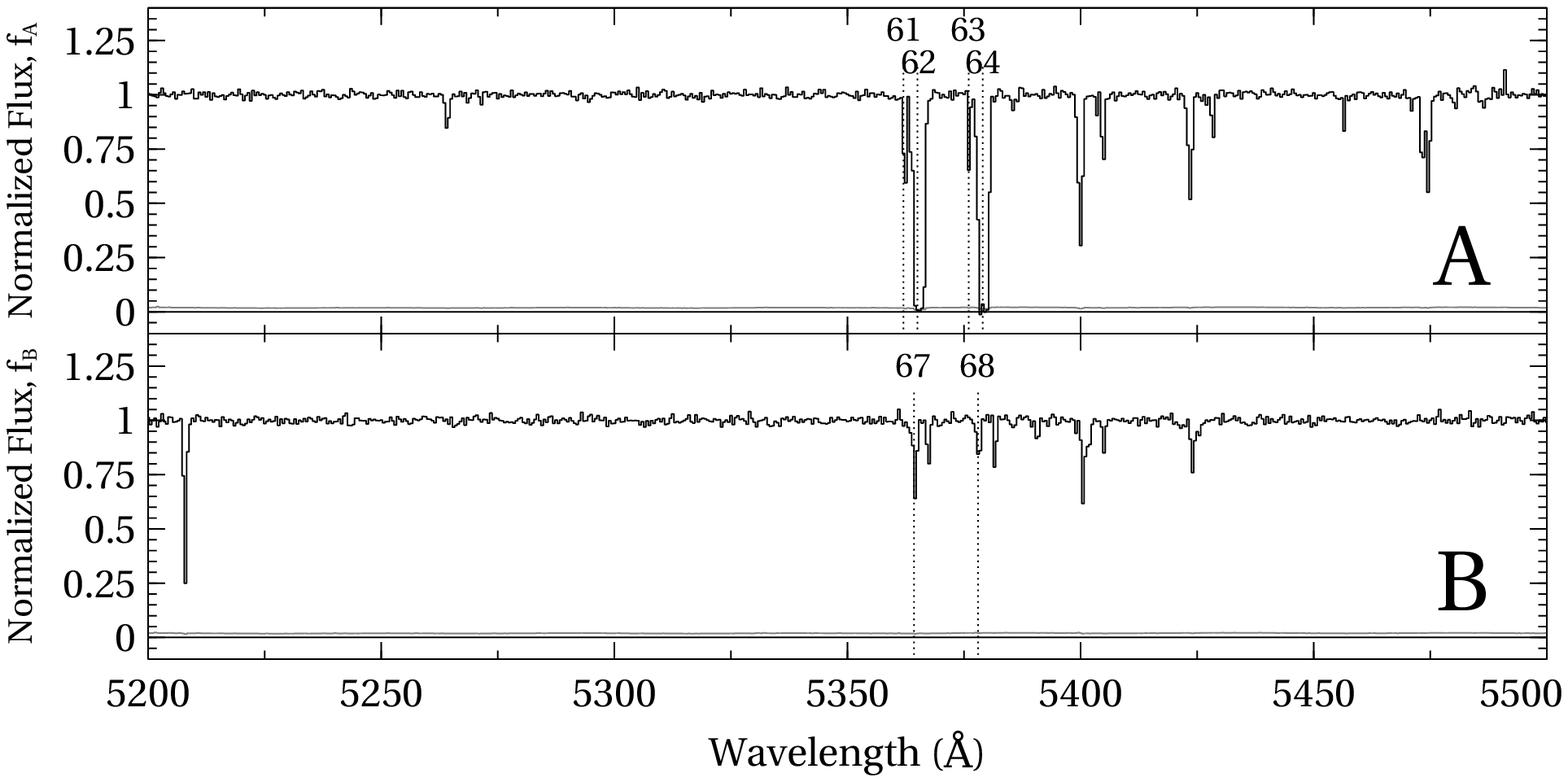}
 \end{center}
 \caption{Continued.}
\end{figure*}

\clearpage
\begin{figure*}
 \begin{center}
  \includegraphics[width=12cm,angle=0]{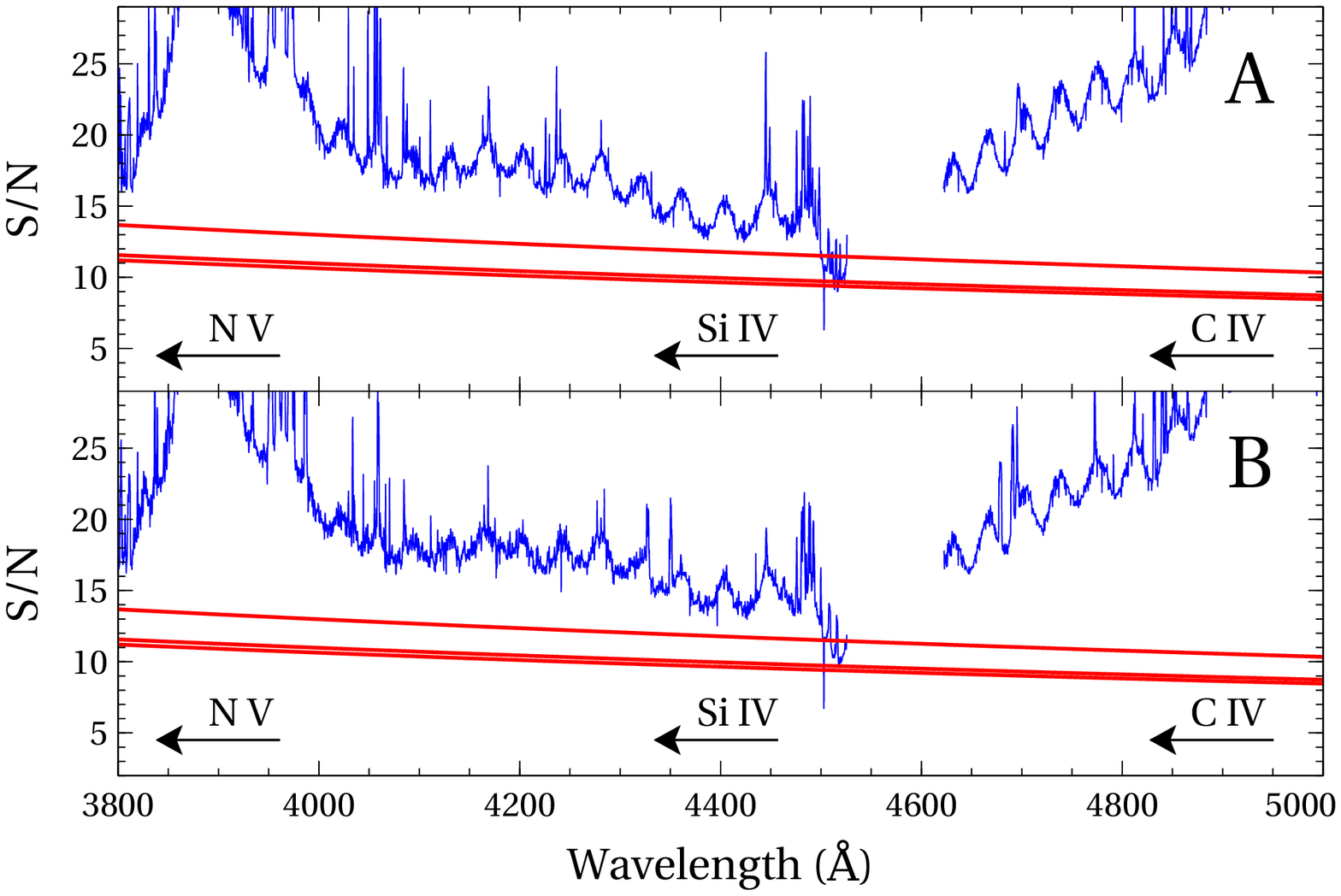}
 \end{center}
 \caption{Signal-to-Noise (S/N) ratio of the image~A and B spectra
   (blue histograms) and the minimum S/N ratio that is necessary for
   detecting homogeneous sample lines for \ion{N}{5}, \ion{C}{4}, and
   \ion{Si}{4}, respectively (red lines from top to bottom; see
   Section 3.1).\label{f2}}
\end{figure*}

\begin{figure*}
\centerline{
\includegraphics[scale=1.0]{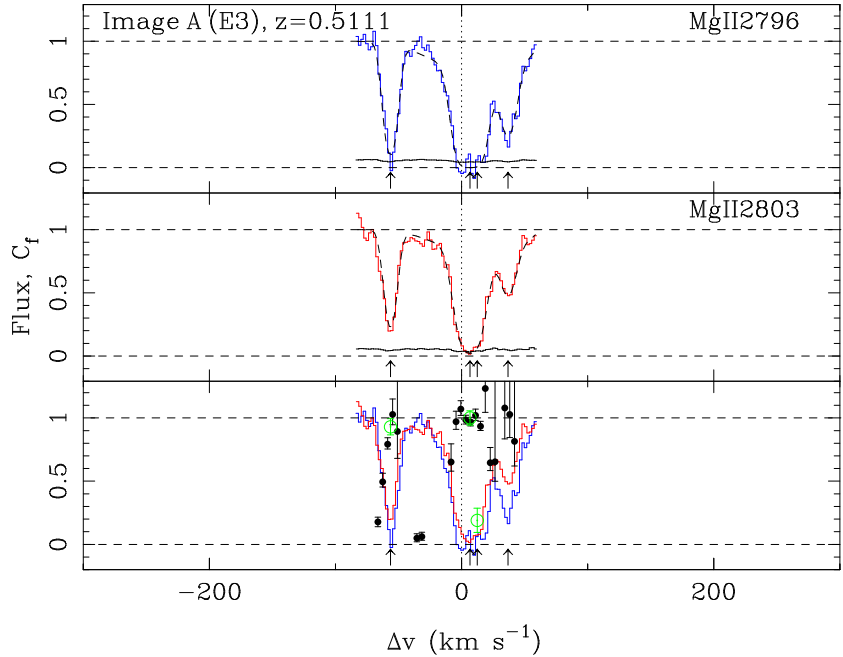}	
}
\caption{Results of partial coverage analysis applied to all
  \ion{C}{4}, \ion{N}{5}, \ion{Si}{4}, and \ion{Mg}{2} NALs detected
  in images~A and B of SDSS~J1029+2623, except for the PALs whose
  results are presented in Figure~\ref{f12}. The horizontal axis
  denotes the relative velocity from the flux-weighted center of the
  system ($\Delta v$) while the vertical axis is the normalized
  flux. The first two panels show the profiles of the blue and red
  members of a doublet (blue and red histograms) with the model
  profile produced by {\sc minfit} superposed (dashed line). The
  positions of the narrow components are marked with upward arrows in
  the bottom of each panel.  The bottom panel shows the covering
  factors with their 1$\sigma$ error bars, measured for each narrow
  component by {\sc minfit} (green circles) or for each pixel (black
  dots). If we had to assume \cf\ = 1 for some components because of
  unphysical \cf\ values derived from {\sc minfit} (see Section 3.2),
  we do not plot them.  \it{The same figures for the other NALs are
    presented in the complete version.}}
\label{f3}
\end{figure*}

\begin{figure*}
 \begin{center}
  \includegraphics[scale=0.8]{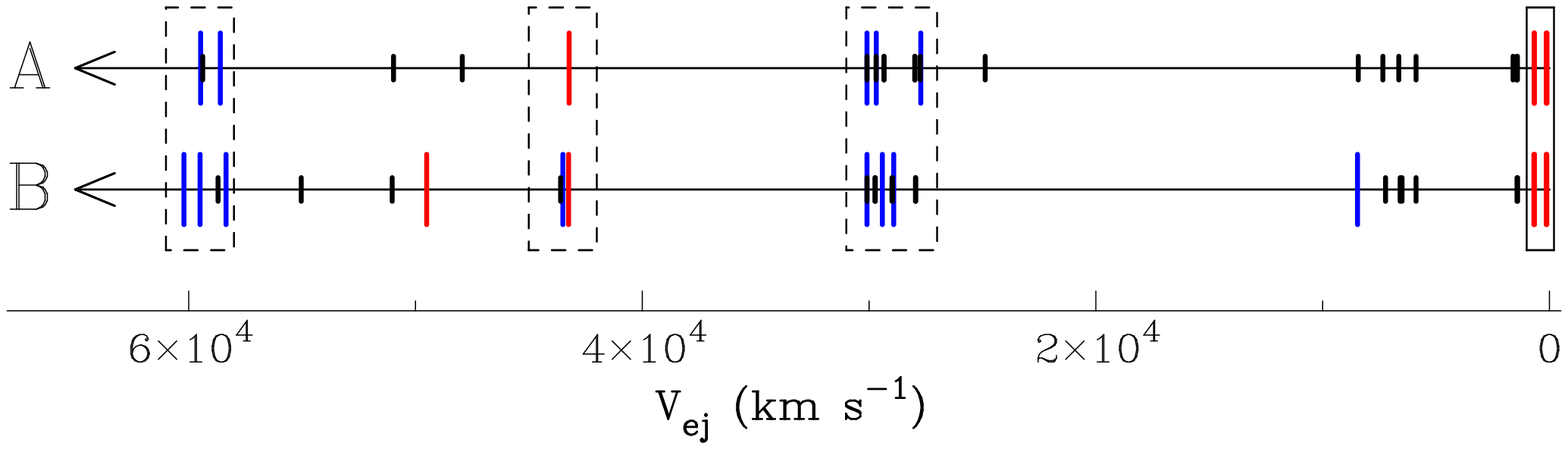}
 \end{center}
 \caption{Ejection velocity distribution of class-A (red line), B
   (blue line), and C (black line) NALs and PALs along the sightlines
   to the images~A and B. All \ion{Mg}{2} NALs are located outside of
   the range of the velocity plot. Each of the red lines at
   \vej\ $\sim$ 0~\kms\ surrounded by a solid rectangle includes both
   \ion{C}{4} and \ion{N}{5} PALs, respectively. Three clustering
   regions are surrounded by dotted rectangles.\label{f4}}
\end{figure*}

\begin{figure*}
 \begin{center}
  \includegraphics[width=12cm,angle=0]{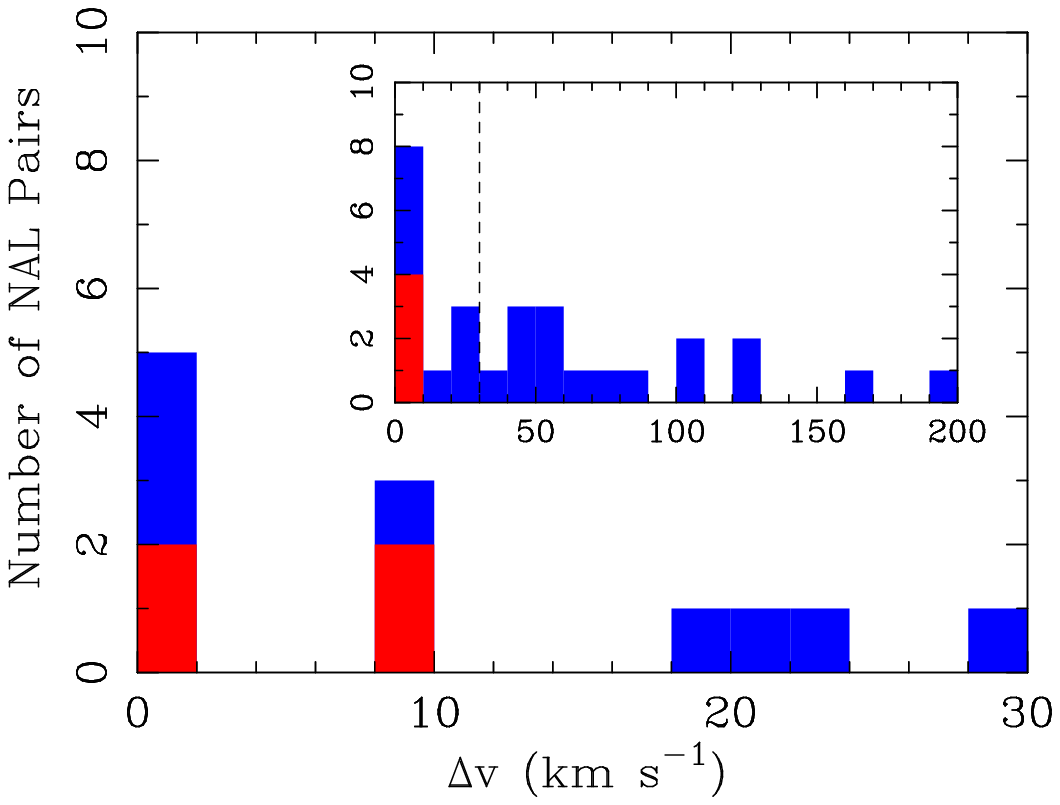}
 \end{center}
 \caption{Distribution of velocity offset ($\Delta v$) between PALs
   (red histogram) and class-A/B NALs (blue histogram) in the two
   different images (image~A and B) that match each other within
   $\Delta v$ $\leq$ 200~\kms, as discussed in Section
   4.1.4.\label{f5}}
\end{figure*}

\begin{figure*}
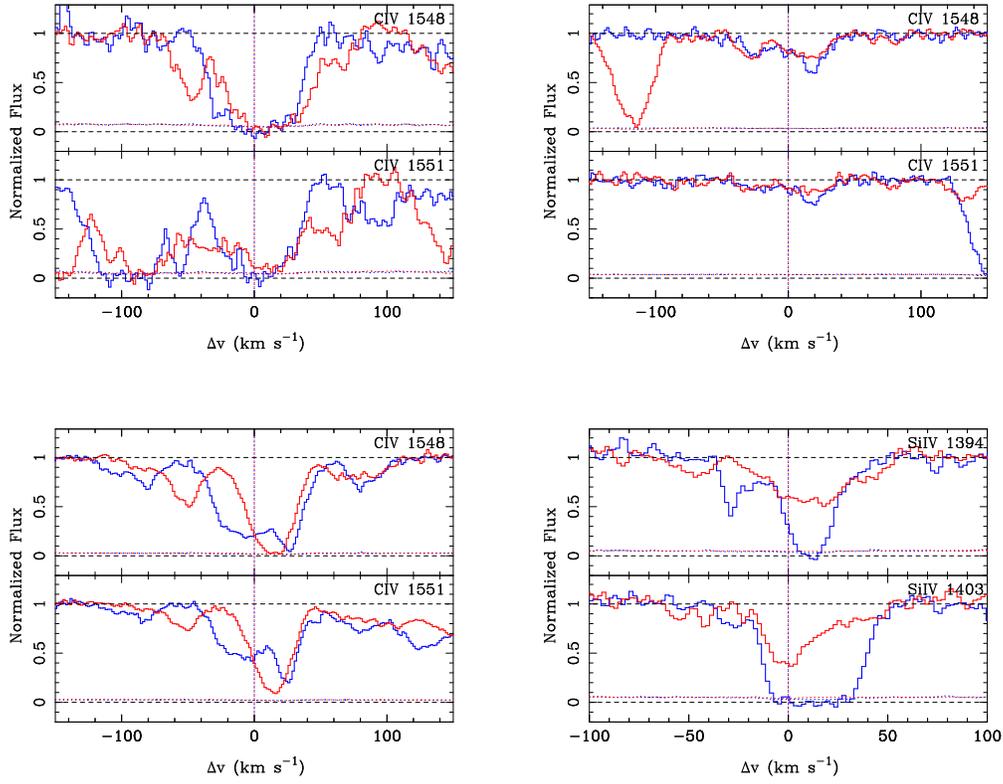

 \begin{center}
  \includegraphics[width=7cm,angle=0]{fig6a.ps}
  \includegraphics[width=7cm,angle=0]{fig6b.ps}
  \includegraphics[width=7cm,angle=0]{fig6c.ps}
  \includegraphics[width=7cm,angle=0]{fig6d.ps}
 \end{center}
 \caption{Comparison of spectra and their 1$\sigma$ errors around
   \ion{C}{4} NALs at \zabs\ $\sim$ 1.8909 (upper left), 2.1349 (upper
   right), 2.1819 (lower left) and \ion{Si}{4} NAL at \zabs\ $\sim$
   1.8909 (lower right) in image~A (blue histogram) and B (red
   histogram) spectra for systems that have a small relative velocity
   between the images.\label{f6}}
\end{figure*}

\begin{figure*}
 \begin{center}
  \includegraphics[width=10cm,angle=0]{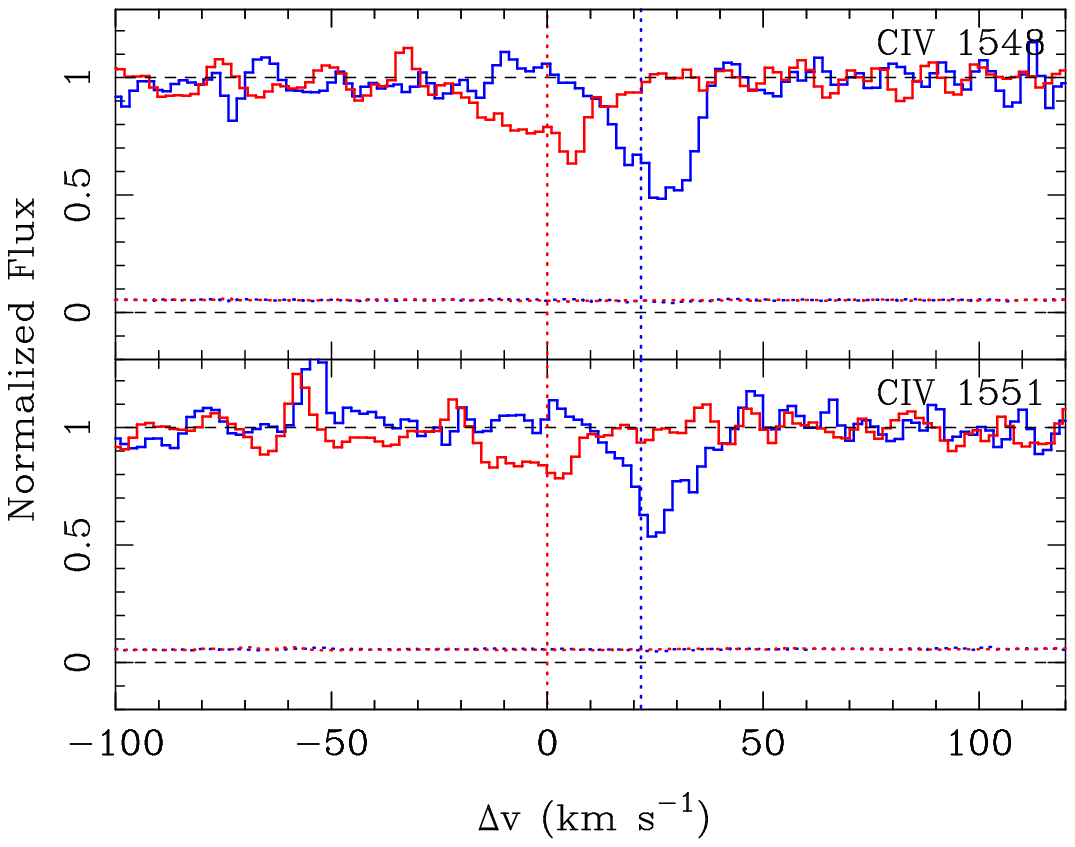}
 \end{center}
 \caption{Same as Figure~\ref{f6} but for class-A \ion{C}{4} NALs at
   \zabs\ $\sim$ 1.7652 in image~A (blue histogram) and at
   \zabs\ $\sim$ 1.7650 in image~B (red histogram). The latter
   corresponds to the system center ($\Delta v$ = 0~\kms) in this
   plot.\label{f7}}
\end{figure*}

\begin{figure*}
 \begin{center}
  \includegraphics[width=8cm,angle=0]{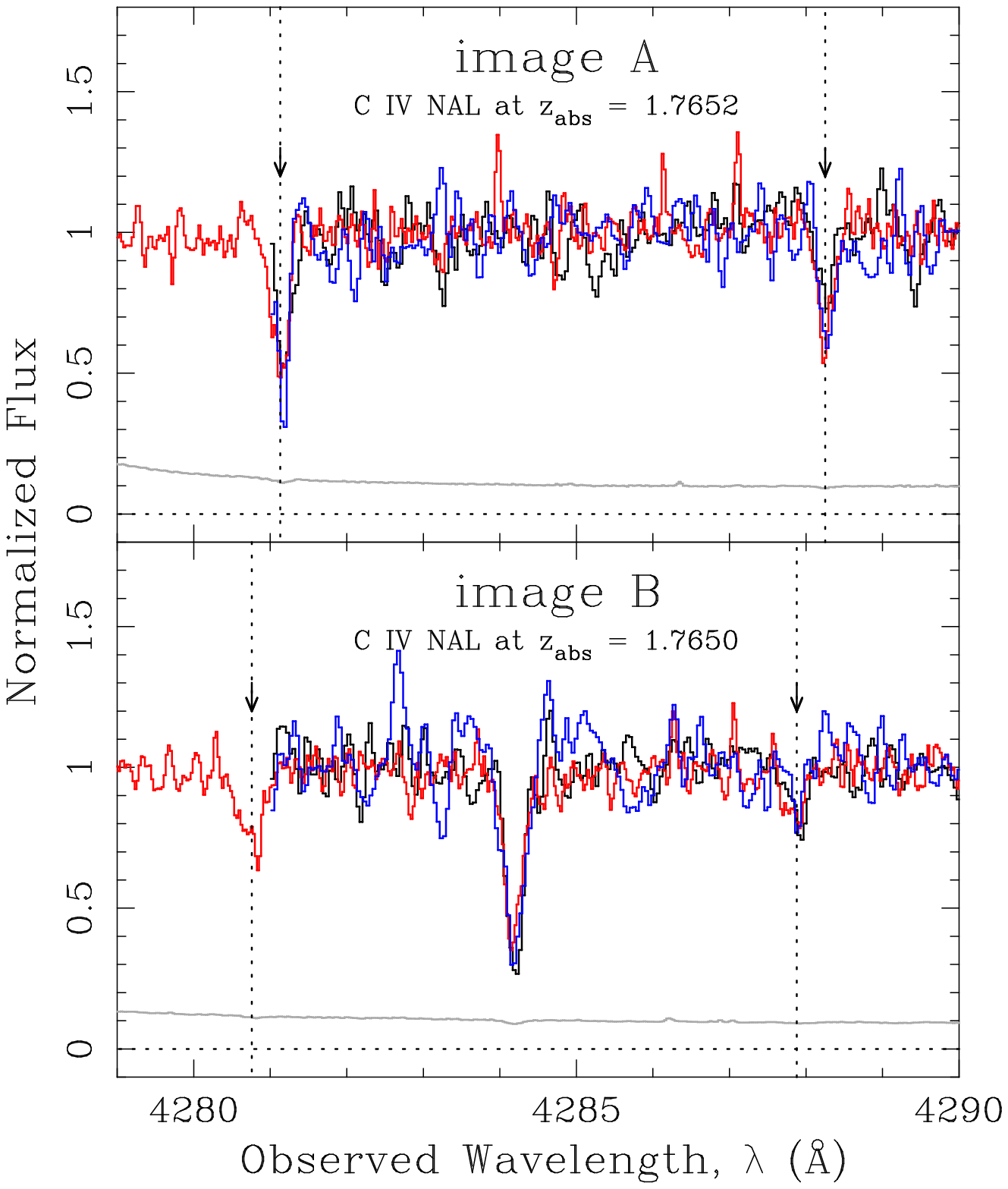}
 \end{center}
 \caption{Comparison of \ion{C}{4} NALs at \zabs\ = 1.7652 in image~A
   (top panel) and at \zabs\ = 1.7650 in image~B (bottom panel).
   Black, red, and blue histograms denote spectra in epoch E1, E2, and
   E3, respectively. The positions of blue and red member of the
   doublet are marked with down arrows and vertical dashed lines.  A
   strong feature at $\lambda$ $\sim$ 4284~\AA\ in the bottom panel is
   \ion{C}{4}~1551 at \zabs\ = 1.7627.  Regions at $\lambda$ $\leq$
   4281~\AA\ are not covered by the Subaru/HDS spectra in epochs E1
   and E3 because they are located at the edge of the CCD.\label{f8}}
\end{figure*}

\begin{figure*}
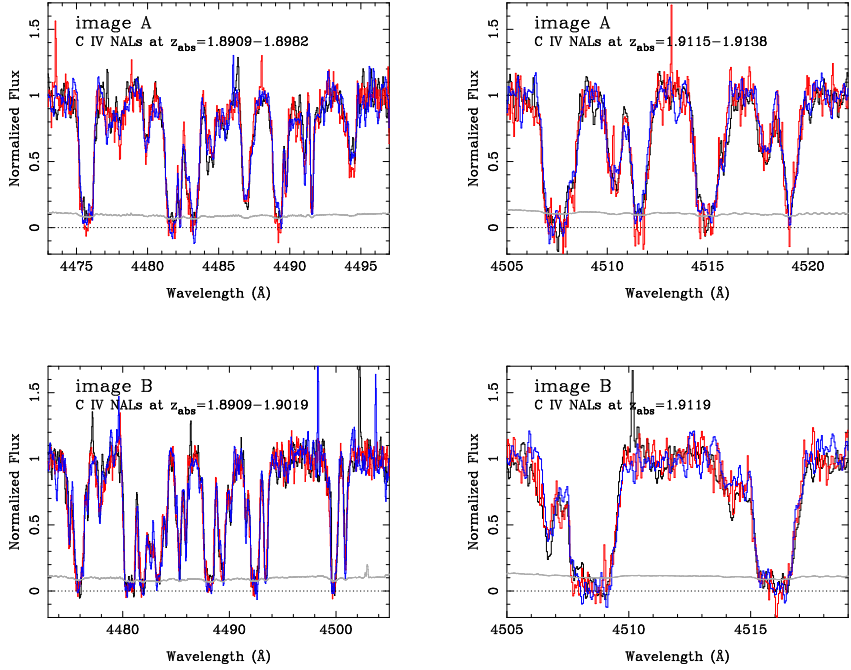

 \begin{center}
  \includegraphics[width=6cm,angle=0]{fig9a.ps}
  \includegraphics[width=6cm,angle=0]{fig9b.ps}
  \includegraphics[width=6cm,angle=0]{fig9c.ps}
  \includegraphics[width=6cm,angle=0]{fig9d.ps}
 \end{center}
 \caption{Comparison of spectra and their averaged 1$\sigma$ errors
   around \ion{C}{4} NALs at \zabs\ $\sim$ 1.8909--1.8982 in image~A
   (upper left), 1.9115--1.9138 in image~A (upper right),
   1.8909--1.9019 in image~B (lower left) and 1.9119 in image~B (lower
   right) taken in epoch E1 (black histogram), E2 (red histogram), and
   E3 (blue histogram).\label{f9}}
\end{figure*}

\begin{figure*}
 \begin{center}
  \includegraphics[width=11cm,angle=270]{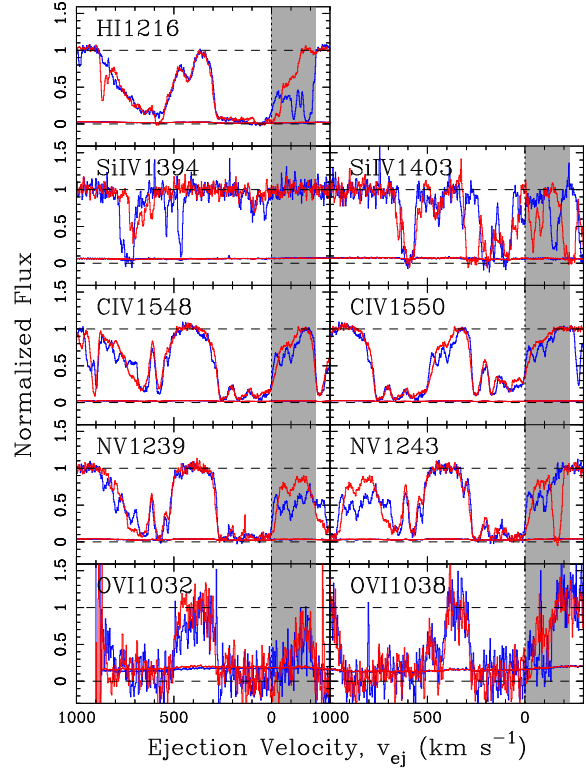}
 \end{center}
 \caption{Comparison of normalized spectra around \lya, \ion{Si}{4},
   \ion{C}{4}, \ion{N}{5}, and \ion{O}{6} PALs detected in the
   images~A (blue) and B (red). The horizontal axis is the offset
   velocity from the quasar emission redshift, and the vertical axis
   is the normalized flux.  The histograms above the zero flux line
   are 1$\sigma$ flux errors.  Due to heavy blending with other lines
   at lower redshift the existence of the \ion{Si}{4} doublet may not
   be present. The \ion{O}{6}~1032 locate at the edge of the observed
   spectrum. Absorption features at \vej\ $<$ 0~\kms\ in the shaded
   areas are narrow PALs, while the other features at \vej\ $>$
   0~\kms\ are broad PALs.\label{f10}}
\end{figure*}

\begin{figure*}
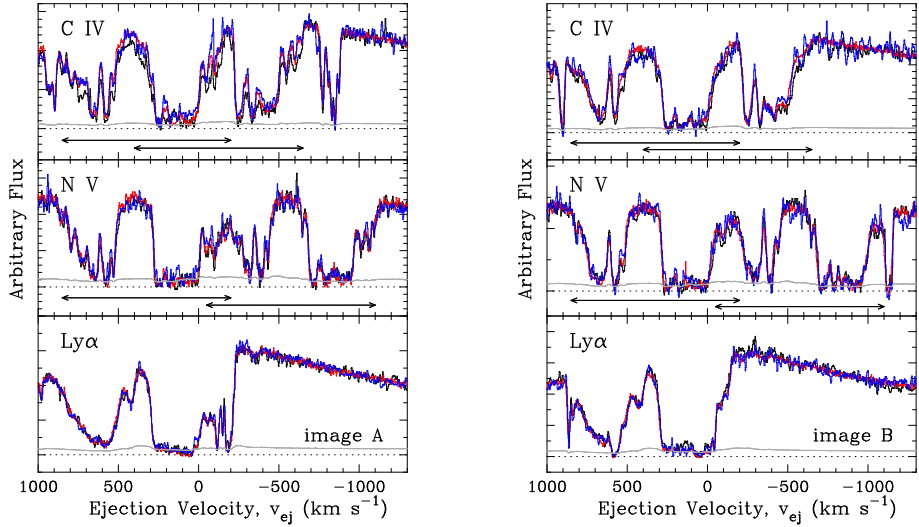

 \begin{center}
  \includegraphics[height=8cm,angle=0]{fig11a.eps}
  \includegraphics[height=8cm,angle=0]{fig11b.eps}
 \end{center}
 \caption{Comparison of \ion{C}{4}, \ion{N}{5}, and \lya\ PALs at
   \zabs\ $\sim$ \zem\ in images~A and B, taken in epochs E1 (black),
   E2 (red), and E3 (blue).  These spectra are scaled to each other by
   applying a least-square method for unabsorbed wavelength regions at
   both sides of the PALs. The vertical axis denotes an arbitrary flux
   because we did not perform absolute flux calibration without flux
   standard stars. The gray histograms above the zero flux line are
   1$\sigma$ flux error after taking averages of those in three
   observing epochs.  Double-headed arrows in the bottom of the first
   two panels mark the velocity ranges of the blue and red members of
   the \ion{C}{4} and \ion{N}{5} doublets.\label{f11}}
\end{figure*}

\begin{figure*}
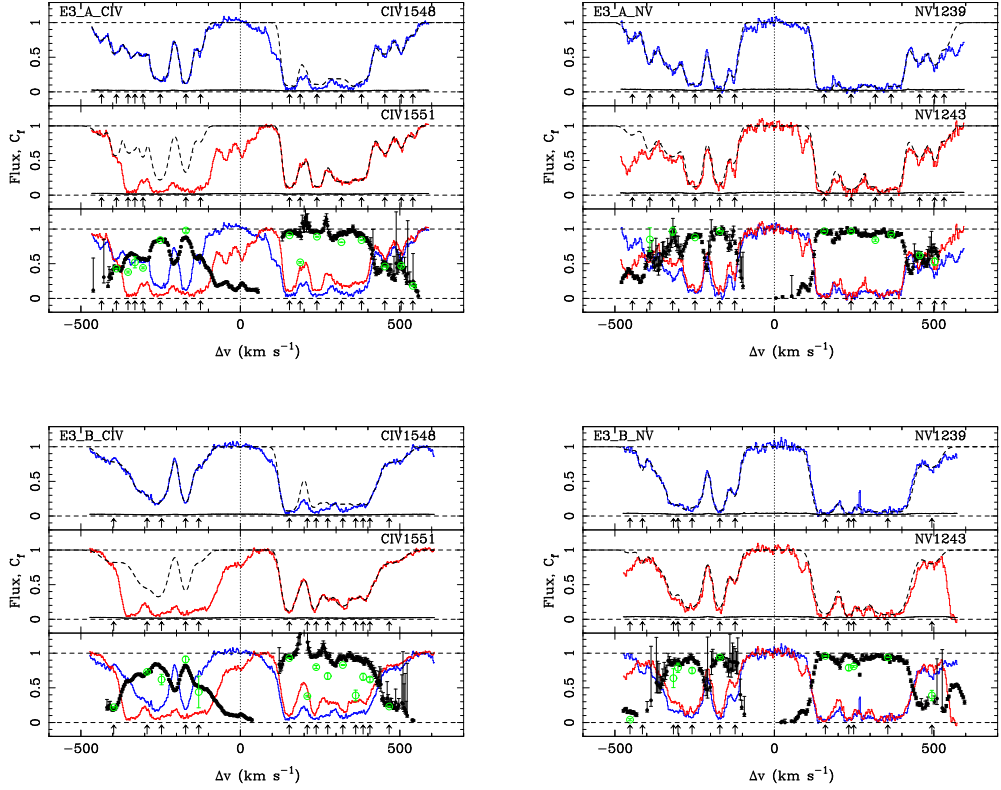

 \begin{center}
  \includegraphics[width=7cm,angle=0]{fig12a.eps}
  \includegraphics[width=7cm,angle=0]{fig12b.eps}
  \includegraphics[width=7cm,angle=0]{fig12c.eps}
  \includegraphics[width=7cm,angle=0]{fig12d.eps}
 \end{center}
 \caption{Same figures as Figure~\ref{f3}, but for \ion{C}{4} and
   \ion{N}{5} PALs at \zabs\ $\sim$ \zem.  We adopt $z$ = 2.1927 as a
   system center only for the purpose of displaying these velocity
   plots.  Because \ion{C}{4} PALs are self-blended, we fit the
   profiles of the blue and red members of the doublet simultaneously
   by multiplying the contributions from them \citep{mis07b}.
   Spectral regions in the top two panels, for which the profiles of
   the blue and red members of a doublet (blue and red histograms) and
   the model profile (dashed line) do not match, suffer from
   self-blending.\label{f12}}
\end{figure*}

\begin{figure*}
 \begin{center}
  \includegraphics[width=13cm,angle=0]{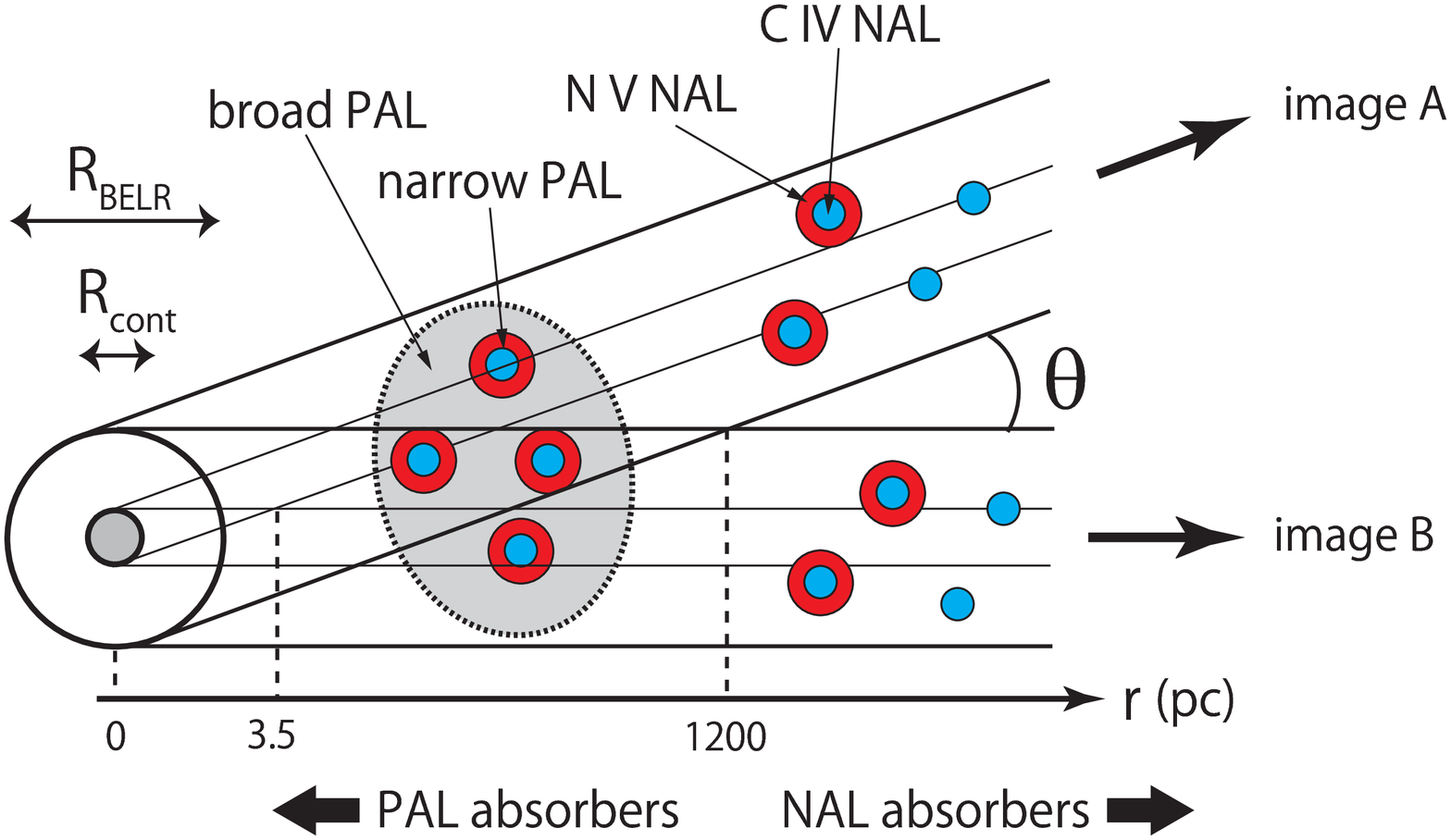}
 \end{center}
 \caption{Possible geometry of the outflowing wind along our two
   sightlines toward the lensed quasar SDSS~J1029+2623, based on our
   observations in three epochs. The gray filled circle corresponds to
   the broad PAL absorber with the minimum acceptable size, while blue
   and red filled circles are \ion{C}{4} and \ion{N}{5} absorbing
   regions in clumpy clouds as candidates for narrow PAL absorber and
   intrinsic NALs. The boundary radius of the continuum source
   (3.5~pc) and BELR (1200~pc) are marked with dashed lines.  The NAL
   absorber's distance is not necessarily larger than 1200~pc,
   although it is at least larger than the distance of the PAL
   absorbers.  We adopt 14.$^{\!\!\prime\prime}$6 as the separation
   angle between two sightlines seen from the source, while
   \citet{mis13,mis14b} used 22.$^{\!\!\prime\prime}$5 assuming it is
   very similar to that seen from us.  See Section 5.3 for
   details.\label{f13}}
\end{figure*}

\end{document}